\documentclass[twocolumn,aps,superscriptaddress]{revtex4-2}
\usepackage{amsmath, amssymb, amsthm, graphicx, epsfig, fancyhdr,epsfig,multirow}
\usepackage[utf8]{inputenc}
\usepackage{amsmath}
\usepackage{amsfonts}
\usepackage{amssymb}
\usepackage{xcolor}
\usepackage{comment}
\usepackage{subcaption}
\usepackage[normalem]{ulem}
\usepackage{tabularx}
\usepackage{comment}
\usepackage[left=2cm,right=2cm,top=2cm,bottom=2cm]{geometry}
\usepackage[font=small,labelfont=bf,textfont=it]{caption}
\usepackage {ulem}

\usepackage{color}

\renewcommand{\sout}{\bgroup \color{red} \ULdepth=-.5ex \ULset}

\begin{document}
\title{Exploring the chiral magnetic effect in isobar collisions through Chiral Anomaly Transport}
\author{Zilin Yuan}
\email{yuanzilin20@mails.ucas.ac.cn}
 \affiliation{School of Nuclear Science and Technology, University of Chinese Academy of Sciences, Beijing, 100049,  P.R. China}
\affiliation{Institue of High Energy Physics, Chinese Academy of Sciences,
Beijing, 100049,  P.R. China}
\author{Anping Huang}
\email{huanganping@ucas.ac.cn}
\affiliation{School of Nuclear Science and Technology, University of Chinese Academy of Sciences, Beijing, 100049,  P.R. China}
\affiliation{School of Material Science and Physics, China University of Mining and Technology, Xuzhou, China}
\author{Guannan Xie}
\email{xieguannan@ucas.ac.cn}
\affiliation{School of Nuclear Science and Technology, University of Chinese Academy of Sciences, Beijing, 100049,  P.R. China}
\author{Wen-Hao Zhou}
\email{zhouwenhao@xaau.edu.cn}
\affiliation{Key Laboratory of Nuclear Physics and Ion-beam Application (MOE), Institute of Modern Physics, Fudan University, Shanghai 200433, China}
\affiliation{Shanghai Research Center for Theoretical Nuclear Physics, NSFC and Fudan University, Shanghai $200438$, China}
\affiliation{Faculty of Science, Xi’an Aeronautical Institute, Xi’an 710077, China}
\author{Guo-Liang Ma}
\email{glma@fudan.edu.cn}
\affiliation{Key Laboratory of Nuclear Physics and Ion-beam Application (MOE), Institute of Modern Physics, Fudan University, Shanghai 200433, China}
\affiliation{Shanghai Research Center for Theoretical Nuclear Physics, NSFC and Fudan University, Shanghai $200438$, China}
\author{Mei Huang}
\email{huangmei@ucas.ac.cn}
\affiliation{School of Nuclear Science and Technology, University of Chinese Academy of Sciences, Beijing, 100049,
  P.R. China}

\begin{abstract}
We investigate the signal of the chiral magnetic effect (CME) in Au+Au collisions and isobar collisions of $_{44}^{96}\text{Ru}+\rm{} _{44}^{96}Ru$ and $_{40}^{96}\text{Zr}+\rm{}_{40}^{96}Zr$ in the newly developed chiral anomaly transport (CAT) module based on the state-of-the-art model a multiphase transport (AMPT). Our numerical simulation results for the ratio charge correlation $\Delta\gamma$ in Ru+Ru and Zr+Zr collisions are close to the latest experimental data. The simulation shows that the CME signal is larger in Ru+Ru collisions than that in Zr+Zr collisions, while the background is smaller, and the upper limit of the CME signal is $15\%$ in the isobar collisions.

\end{abstract}
\maketitle

\section{introduction}
When two ion nuclei collide in relativistic heavy ion collisions, a hot/ dense, and deconfined state of nuclear matter known as Quark-Gluon Plasma (QGP) is formed. This state has attracted significant attention, aimed at investigating the properties of QGP as described by quantum chromodynamics (QCD) at high energy densities. For non-central heavy ion collisions, the magnetic field with the strength of $eB\sim10^{18-20}\ \text{Gauss}$ and life time of $10^{-22}$ second can be generated \cite{Skokov:2009qp,Deng:2012pc}, and the orbital angular momentum with the magnitude $\mathcal{O}(\text{10}^4-\text{10}^5) \hbar$ \cite{Becattini:2007sr} corresponding to the local angular velocity 0.01 to 0.1 GeV \cite{Li:2017slc} can be produced. QCD matter under external magnetic field shows quite a few non-trivial phenomena, including the manifestation of the Chiral Magnetic Effect (CME)~\cite{Kharzeev:2007tn,Kharzeev:2007jp,Fukushima:2008xe}, and chiral vortical effect (CVE) \cite{Kharzeev:2010gr}. 

During the collision, the magnetic field rapidly intensifies until the nuclei collide, reaching its maximum strength in the narrow collision region influenced by the spectators. After this maximum strength, the contribution from the receding spectators starts to decrease, while the remnants significantly slow the decay of the magnetic field. The magnetic field's evolution within the Quark-Gluon Plasma (QGP) demonstrates a time-dependent behavior
\cite{Deng:2012pc,McLerran:2013hla}: $eB(t,r)=\frac{eB(0,r)}{1+(\frac{t}{\tau_{B}})^{2}}$, where the $eB(0,r)$ is the initial magnetic field at $t=0$ from spectators and the $\tau_{B}$ is the lifetime of the magnetic field. 

The CME is related to the nontrivial topological gauge field with the integer Chern-Simons number $Q_W$, 
\begin{equation}
Q_w
= \frac{g^2}{32\pi^2} \int \mathrm{d}^4 x\, 
F_{\mu \nu}^a \tilde F^{\mu\nu}_a
\in \mathbb{Z},
\end{equation}
where $g$ is the QCD coupling constant,  $\mathrm{tr}\, t_a t_b = \delta_{ab} / 2$ are the normalized generators, and
$F_{\mu\nu}^a$ and $\tilde F^a_{\mu\nu} = \tfrac{1}{2} \epsilon_{\mu\nu}^{\phantom{\mu\nu}\rho\sigma} F^a_{\rho \sigma}$ are the gluonic field tensor and its dual, respectively.
The topological configurations could break the parity (P) and charge-parity (CP) symmetry of QCD \cite{Chern:1974ft,Jackiw:1976pf}.
The quarks will obtain a chirality imbalance $N_{5}=2N_{f}Q_{W}$ in the QGP when the parity and charge-parity symmetries are violated. Under these conditions, there would be a charge current developed along the direction of the magnetic field:
\begin{equation}
    \mathbf{J}_{V}=\frac{Q^{2}\mu_{5}}{2\pi^{2}}\mathbf{B}, \label{current and mu5}
\end{equation}
where $\mu_{5}$ is the chiral chemical potential caused by the chirality imbalance of right handed and left handed quarks. This phenomena is called the Chiral Magnetic Effect \cite{Fukushima:2008xe}. The Chiral Magnetic Effect will induce a separation of quarks with positive and negative charges along the direction of the magnetic field. Consequently, this leads to a final charge number difference that is perpendicular to the reaction plane.To quantify the charge separation induced by the CME, the signal is phenomenologically characterized by sine terms in the Fourier decomposition of the azimuthal distribution of charged particles, as follows\cite{Voloshin:2004vk,Masera:2009zz,Kharzeev:2004ey}:
\begin{equation}
\label{particle distribution}
\begin{split}
  \frac{\mathrm{d}N_i}{\mathrm{d}\phi_i}&=\frac{N_i}{2\pi}[1+2a_1 \sin(\phi_i-\psi) \\ &+\sum_{k}2v_k\cos(k(\phi_i-\psi))]
\end{split}
\end{equation}
where $N_i$ is the total multiplicity of a given type particle, $v_n$ are the flow coefficients, $a_1$ is the coefficient of local CP violation and $\Psi_{\text{RP}}$ is the azimuthal angle of the reaction plane. Owing to the Chiral Magnetic Effect (CME), the chiral anomaly present in the quark-gluon plasma stage of heavy ion collisions is transformed into an observable charge-dependent azimuthal anisotropy, denoted as
$a_1$ at the final freeze-out stage \cite{Kharzeev:2015znc}. However, this effect may vary from event to event, since the direction of the current perpendicular to the reaction plane is indeterminate for each collision event. To address this challenge, the key observable measurement used to detect the CME is the two-particle correlation\cite{Voloshin:2004vk},
\begin{equation}\label{eq:gamma}
\begin{split}
\gamma & = \left\langle\cos(\phi_{\alpha} +\phi_{\beta} -2\Psi_{\text{RP}} )\right\rangle \\
    & =\left\langle\cos(\phi_{\alpha} -\Psi_{\text{RP}} )\cos(\phi_{\beta} -\Psi_{\text{RP}} )\right\rangle\\ &-\left\langle\sin(\phi_{\alpha} -\Psi_{\text{RP}} )\sin(\phi_{\beta} -\Psi_{\text{RP}} )\right\rangle \\
    & =(\left\langle v_{1,\alpha}v_{2,\beta} \right\rangle + B_\mathrm{IN})-(\left\langle a_{1,\alpha}a_{2,\beta} \right\rangle + B_\mathrm{OUT}),
    \end{split}
\end{equation}
We reconstruct the event plane to estimate the reaction plane as the method used in Refs. \cite{STAR:2009tro,STAR:2004jwm,CMS:2016wfo,Wang:2022eoo},
\begin{equation}
\Psi_{\text{EP}}=(\tan^{-1}(\frac{\sum_{i}\omega_i \sin(2\phi_i)}{\sum_{i} \omega_i\cos(2\phi_i)}))/2.
\end{equation}
 $\phi_{\alpha}$ and $\phi_{\beta}$ in 
Eq.\eqref{eq:gamma} are the azimuthal angles of the produced charged particles $\alpha$ and $\beta$, respectively, with $\alpha$ and $\beta$ denoting either positive or negative charges. Here $\omega_i$ is the weight for the $i$th particle, which is chosen to be linear with $p_T$ up to 2 GeV/$c$ and then constant at a value of 2 for higher momenta. While $ \left\langle v_{1,\alpha}v_{2,\beta} \right\rangle$ and $\left\langle a_{1,\alpha}a_{2,\beta} \right\rangle$ in the fourth line in Eq.\eqref{eq:gamma} represent projected azimuthal correlation in the in-plane and out-of-plane. The $\left\langle a_{1,\alpha}a_{2,\beta} \right\rangle$ term serves as an evidence for charge separation perpendicular to the reaction plane. The $ \left\langle v_{1,\alpha}v_{2,\beta} \right\rangle$ term is expected to be negligible, because the average directed flow $\left\langle v_{1}\right\rangle$ should be zero over the detected rapidity region due to symmetry. Additionally, $B_\mathrm{IN}$ and $B_\mathrm{OUT}$ denote the backgrounds of the in-plane and out-of-plane correlations, respectively. The primary contribution to this background arises from elliptic flow and particle multiplicity, which is influenced by resonance decays, local charge conservation, and transverse momentum conservation\cite{Zhao:2019hta,Wang:2016iov,Schlichting:2010qia,Voloshin:2004vk}. For example, if particles $\alpha$ and $\beta$ are the products of a resonance decays, their correlations include a contribution given by $\cos(\phi_{\alpha} +\phi_{\beta} -2\Psi_{\text{RP}} )\approx \cos(\phi_{\alpha} +\phi_{\beta} -2\phi_{res} ) v_{2}$ where $v_{2}$ represents the elliptic flow of the resonance and $\phi_{res}$ is the azimuthal angle of the resonance. Overall, the background of this correlation can be expressed as follows:
\begin{equation}
    \Delta\gamma_{bkg}=\frac{N_{2p}}{N^{2}}\left\langle\cos(\phi_{\alpha} +\phi_{\beta} -2\phi_{2p} )\right\rangle v_{2,2p},
\end{equation}
where $N$ is the multiplicity of a single charge, $N_{2p}$ is the number of correlated pairs such as from parent resonances, $\phi_{2p}$ and $v_{2,2p}$ is the azimuthal angle and elliptic flow of the parent respectively. 

If the charge distribution is symmetric about the reaction plane, it is expected as $\gamma_{++}=\gamma_{--}=-\gamma_{+-}>0$, but the experiment gives $\gamma_{++}=\gamma_{--}\gg-\gamma_{+-}$. The confusing result maybe due to particle interaction with the medium \cite{Kharzeev:2007jp}.
The rest sign-independent background could be eliminated mostly by taking a difference between opposite-sign and same-sign $\gamma$:
\begin{equation}
    \Delta\gamma=\gamma_{os}-\gamma_{ss}=\Delta\gamma_{sig}+\Delta\gamma_{bkg}.
\end{equation}

It was proposed in \cite{Voloshin:2010ut,Deng:2016knn} that the isobaric collisions of $_{44}^{96}\text{Ru}+\rm{} _{44}^{96}Ru$ and $_{40}^{96}\text{Zr}+\rm{}_{40}^{96}Zr$  may provide an ideal tool to separate the CME signal from the large
 background. Recently, the STAR Collaboration has released the results of a blind analysis from isobar collisions\cite{STAR:2021mii}. However, the results from the STAR analysis were inconsistent with the predictions in \cite{Voloshin:2010ut,Deng:2016knn}.

In order to explain the CME signal and the background, the reaction plane independent correlation
\begin{equation}
\begin{split}
   &\delta=\left\langle\cos(\phi_{\alpha} - \phi_{\beta})\right\rangle\\
   &\Delta\delta= \delta_{os}- \delta_{ss}
\end{split}
\end{equation}
should be taken into account as well. This two-particle correlation is proportional to $\left\langle a_{1,\alpha}a_{2,\beta} \right\rangle$ but dominated by the elliptic flow of resonance decay that contributes to the interpretation of background correlations. We can obtain the $(\left\langle v_{\alpha}v_{\beta} \right\rangle - B_\mathrm{IN})=(\gamma+\delta)/2$ term and $(\left\langle a_{\alpha}a_{\beta} \right\rangle - B_\mathrm{OUT})=(\gamma-\delta)/2$ term individually, which also show interesting results \cite{Bzdak:2009fc}. 

The starting point of this work is that the data given by the STAR detector experimental collaboration is inconsistent with the results of theoretical simulations such as UrQMD and HIJING \cite{STAR:2009tro}. Due to the absence of cooperative effect of chiral anomaly and magnetic field, these models cannot describe the chiral magnetic effect even though they can well describe other physical quantities. Considering the large magnetic field presents at the early stage of collision, it is necessary to develop a model to study the Quark-Gluon plasma evolution under magnetic field. We develop such a model based on the AMPT model. 

The AMPT model has the advantage to simulate the parton evolution, and it consists of four parts: HIJING, ZPC, quark coalescence process and ART\cite{Lin:2004en}.
The HIJING part simulates the initial condition of nucleon collision for a certain impact parameter. The ZPC part is a transport model of parton cascade. This part only includes the uniform linear motion and elastic collision process. The quark coalescence part makes the freezeout partons combine to hadrons by minimizing the distances between quarks of hadrons in the coordinate space, and the ART part is a hadron relativistic transport model. At the end of AMPT, it outputs the final particle information after the last collisions of hadrons at a cutoff time.

Same as the previous work \cite{Ma:2011uma}, we choose the AMPT model to simulate the evolution of QGP in strong electromagnetic field. Instead of a given initial charge separation as before, the initial chirality imbalance is calculated self-consistetly to initialize the helicity assigned to each particle,  in this sense the CME signal comes from parton cascade in electromagnetic field rather than an initial charge separation effect. In this work, the parton cascade part of AMPT model dose not have the ability to deal with complex dynamic of a chiral system in response to an external magnetic field by adopting the chiral kinetic theory. We update a chiral anomaly transport (CAT) module based on AMPT model to study the charge separation. In section \ref{AMPT model and the change}, we make a brief description of AMPT model and discuss the improvements of CAT module.  In section \ref{isobarcollision}, we search the detail of background and signal in the two-particle correlation from AMPT-CAT and then observe the signal of CME in isobar collisions at 200 GeV.

\section{The chiral anomaly transport module}\label{AMPT model and the change}

The chiral anomaly transport module is built upon A Multi-Phase Transport model (AMPT) with string melting mechanism\cite{Lin:2004en}, and to focus on solving the Boltzmann equation of quarks which include the chirality imbalance and the magnetic field. The initial particle distribution in coordinate space and momentum space are generated by HIJING as the initial conditions of CAT, from which the magentic field and chiral anomaly are calculated. 
The main component of CAT is a new parton cascade program, which replaces the ZPC module in the original AMPT, using chiral kinetic theory to dynamically generate charge separation along the magentic field with chirality imbalance. After parton cascade, the charge separation of freeze out quarks transform to the charge separation of hadrons during quark coalescence and hadron evolution.

Following Ref \cite{Deng:2012pc}, the initial electromagnetic field from spectator protons in HIJING at $t_{0}=0\ \text{fm}/c$ is evaluated according to 

\begin{equation}\label{mg12}
\begin{split}
e\mathbf{E}(0,r)=\alpha_{EM}\sum_{n} Z_{n}
\frac{\mathbf{R}_{n}(1-{v}_{n}^{2})}{({R}_{n}^2-[\mathbf{v}_{n}\times \mathbf{R}_{n}]^{2})^{3/2} },\\
e\mathbf{B}(0,r)=\alpha_{EM}\sum_{n} Z_{n}
\frac{\mathbf{v}_{n}\times \mathbf{R}_{n}(1-{v}_{n}^{2})}{({R}_{n}^2-[\mathbf{v}_{n}\times \mathbf{R}_{n}]^{2})^{3/2} },
\end{split}
\end{equation}
where $\mathbf{R}_{n}=\mathbf{r}-\mathbf{r}'_{n}$ is the relative position vector from a field point $\mathbf{r}$ to a source point $\mathbf{r}'_{n}$ at the initial time $t_{0}=0 $ fm/c, and $\alpha_{EM}$ is the
EM fine-structure constant, defined as $\alpha_{EM} = {e^2}/{4\pi}\approx 1/137$. 
The time evolution of magnetic field is approximated as follows \cite{McLerran:2013hla,Deng:2012pc}

\begin{equation}
    \label{evolution of mg}
    e\mathbf{B}(t,r)=\frac{e\mathbf{B}(0,r)}{1+(t/\tau)^{2}}
\end{equation}
    where $\tau = 0.4$ fm/c is the lifetime of magnetic field.
    
\begin{figure}[!ht]
\hspace*{-6 mm}
\includegraphics[scale=.4]{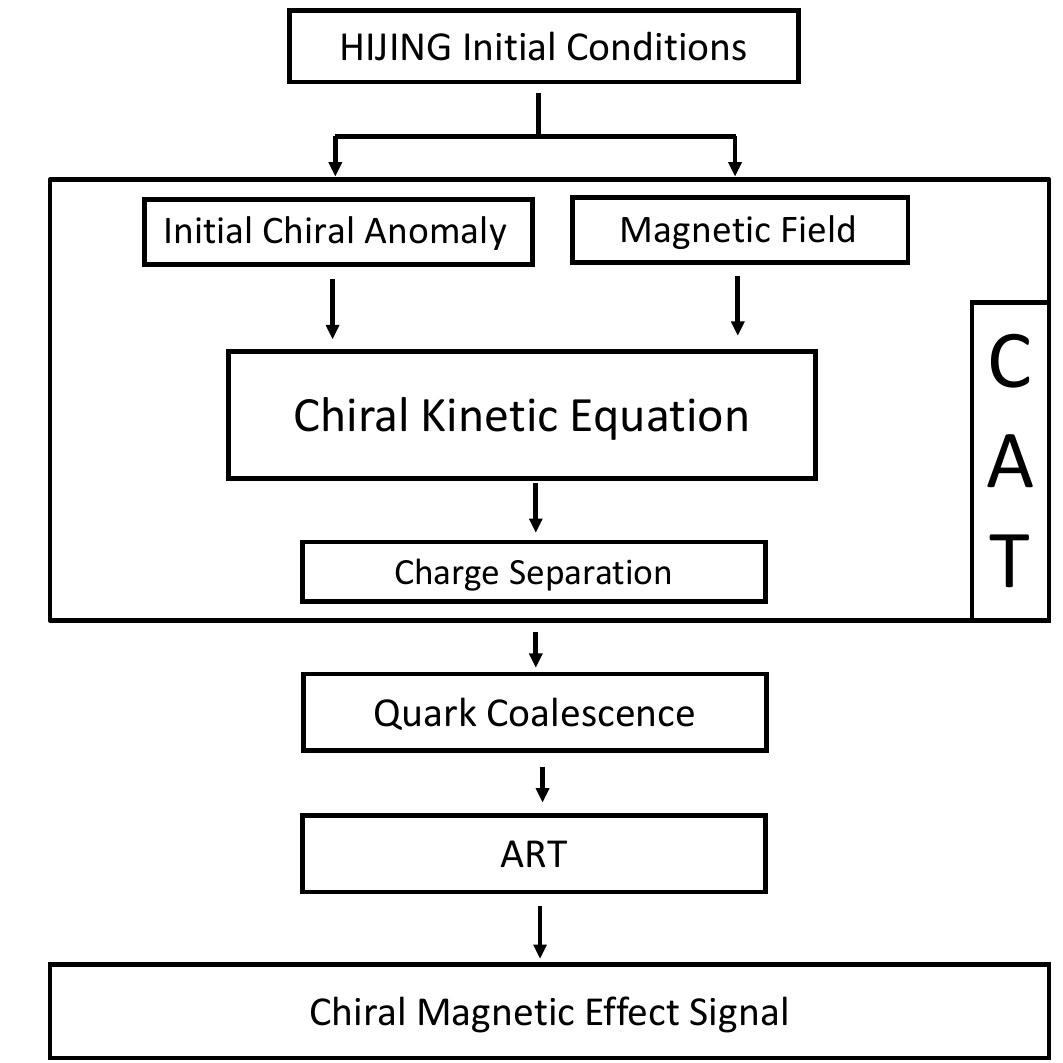}
\caption{(Color online) Illustration of the structure of the chiral anomaly transport module based on AMPT model. The initial conditions of nucleons and quarks are loaded from HIJING. The core component of CAT corresponds to the parton cascade program encompassing the magnetic field, the axial charge and the chiral kinetic theory. The CME observable is calculated by the final hadrons after hadron rescattering.}
\label{CAT-Structure}
\end{figure}

The initial local chirality density for each space cell is defined as $n_{5}  =\mu_{5}^{3}/(3\pi^{2})+(\mu_{5}T^{2})/3$, with  chiral chemical potential $\mu_{5}$ and the local effective temperature $T$ \cite{Lin:2014tya}. $\mu_{5}$ generated by the sphaleron transition rate $\mu_{5}=\pm\sqrt{3\pi}\sqrt{\left(320N_{f}^{2}\Gamma_{ss}/T^{2}-T^{2}/3\right)}$ \cite{Chao:2013qpa} can be either positive or negative for different events, but the positivity or negativity remains consistent in each cell within the same event. In the collisions at 200 GeV, we take 
\begin{equation}
\label{mu5eq}
\mu_5 = \pm(2.1T + \sqrt{eB})\,\text{GeV},
\end{equation}
which can well describe the CME in the Au+Au collision at 200GeV in the previous work \cite{Yuan:2023skl}.
 It should be noticed that averaged values of $\mu_{5}$ and $n_{5}$ calculated event-by-event are zero. 
 For convenience, we redefine the $\mu_{5}$ and $n_{5}$ as  $\mu_{5}=\sqrt{\langle\mu^{2}_{5}\rangle_\text{event}}$ and
 $n_{5}=\sqrt{\langle n^{2}_{5}\rangle_\text{event}}$.

 The chiral kinetic theory, aimed to solve the kinetic equation which takes the following form, for massless chiral particles in magnetic field when $|\mathbf{p}|>eB$\cite{Son:2012wh, Stephanov:2012ki, Huang:2018wdl, Zhou:2018rkh,Sun:2018idn}:
 \begin{eqnarray} \label{ckt}
 \left[ \partial_t + {\dot{\bf{x}}}\cdot   {\nabla}_{x}  +  {\dot{\bf{p}}} \cdot \nabla_{p} \right] f_i({\bf{x}},{\bf{p}},t) = C[f_i]  \,\, ,
 \end{eqnarray}
 \begin{eqnarray} \label{eom}
\sqrt{G}\dot{\mathbf{x}}  =\mathbf{\hat{p}} +q_{i} \mathbf{B}(\mathbf{\hat{p}} \cdot \mathbf{b}),\,\, \sqrt{G}\dot{\mathbf{p}}  =q_{i}\mathbf{\hat{p}}\times \mathbf{B}.
\end{eqnarray}
where $f_{i}$ denotes the distribution function of a specific quark $i$, $q_{i}$ stands for the charge of the quark $i$, $\mathbf{B}$ is the magnetic field, and $\mathbf{b}=\chi \frac{\mathbf{p}}{2|\mathbf{p}|^{3}}$ represents the Berry curvature, where $ \chi = \pm 1$ denotes the helicity. The factor $\sqrt{G}=1+\chi \frac{\hbar \mathbf{B} \cdot \mathbf{p}}{2|\mathbf{p}|^{3}}$. In this model, we employ a massless approximation for quarks and treat the helicity as a good quantum number but modify $|\mathbf{p}|$  to $\mathbf{E_{p}}=\sqrt{|\mathbf{p}|^2+m^2} $, because of the finite but small mass of quark \cite{Chen:2013iga}.

We summarize the main difference between AMPT with ZPC module and AMPT with CAT module in TABLE \ref{cat_AMPT}.
\begin{widetext}
\begin{center}
\begin{table}[]
    \centering
    \begin{tabular}{c|c|c}
            & ZPC module &  CAT module\\
                  \hline
    magnetic field  &  no         & $e\overrightarrow{B}_{0}(r)=\alpha_{EM}\sum_{n} Z_{n}\frac{\overrightarrow{v}_{n}\times \overrightarrow{R}_{n}(1-\overrightarrow{v}_{n}^{2})}{(\overrightarrow{R}_{n}^2-[\overrightarrow{v}_{n}\times \overrightarrow{R}_{n}^{2}])^{3/2} } $ \\
          \hline
    chiral charge  &   no        & $ \mu_5 = \pm(2.1T + \sqrt{eB})$ \\
          \hline
    vorticity field &  no         & $\omega(r)=\gamma^2 \nabla\times v(r)$ \\
          \hline
    kinetic equation &  $\dot{\mathbf{x}}=\mathbf{p}/E$         & $\sqrt{G}\dot{\mathbf{x}}  =\mathbf{\hat{p}} +q_{i} \mathbf{B}(\mathbf{\hat{p}} \cdot \mathbf{b})$ \\
            &  $\dot{\mathbf{p}}=0$         & $\sqrt{G}\dot{\mathbf{p}}  =q_{i}\mathbf{\hat{p}}\times \mathbf{B} $ \\
                  \hline
    CME signal &   input at begining  & from evolution \\
          \hline
    mean field &   no        & generated each time-step \\
    \end{tabular}
    \caption{The comparison between AMPT with ZPC module and AMPT with CAT module. }
    \label{cat_AMPT}
\end{table}
\end{center}
\end{widetext}

\begin{figure}[!ht]
\hspace*{-6 mm}
\includegraphics[scale=0.38]{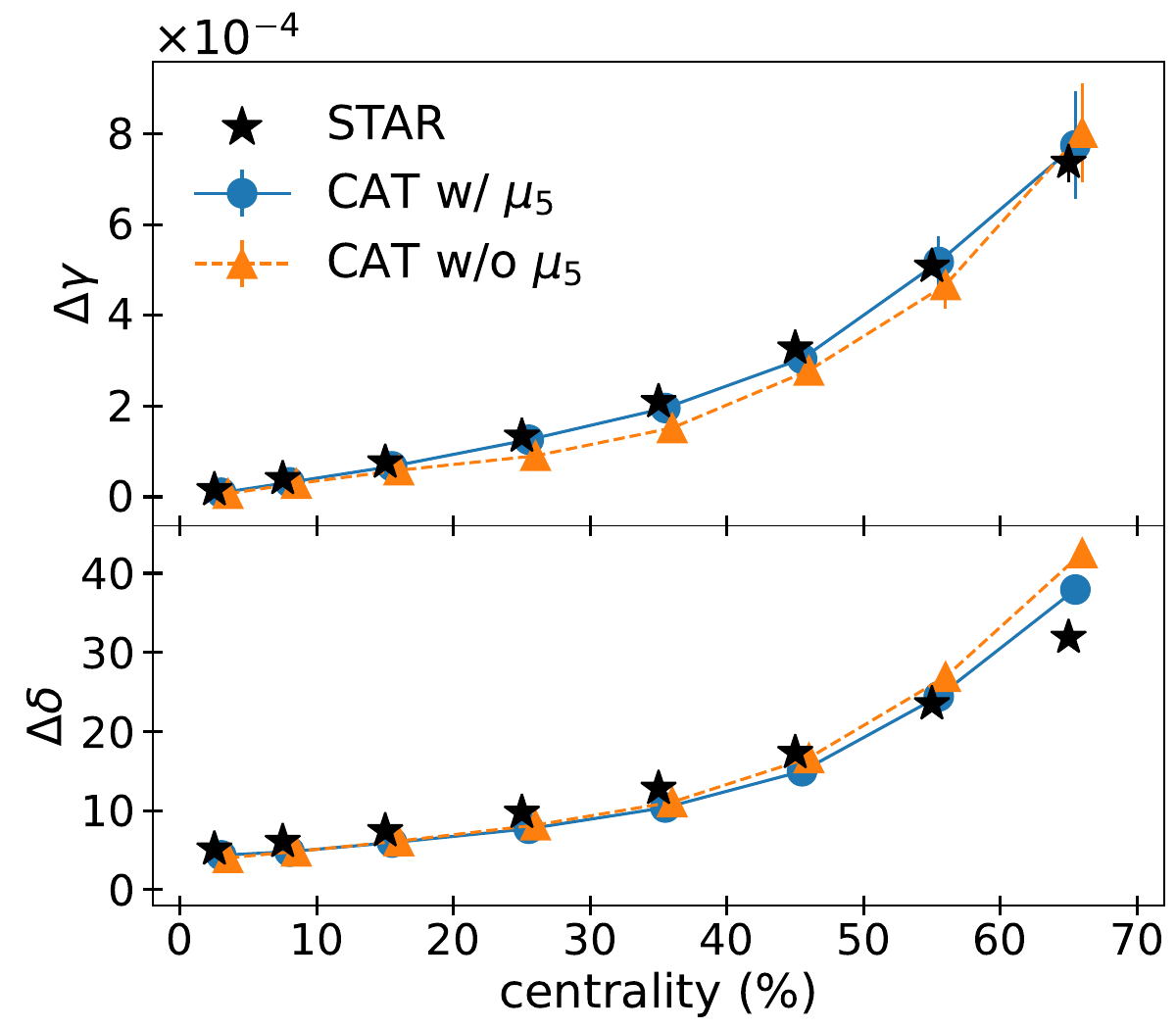}
\caption{The CAT results on the centrality dependences of the correlations  $\Delta \gamma$ (upper panel) and $\Delta \delta$(lower panel) in Au+Au collisions at 200 GeV, compared with the STAR data (stars). The CAT results are calculated by the CAT with or without chiral chemical potential $\mu_5$, shown by blue and orange curves, respectively.}
\label{auauresult}
\end{figure}

In the Au+Au collisions at 200 GeV, the chiral anomaly transport program can well describe the STAR data as shown in \cite{Yuan:2023skl} and Fig.\ref{auauresult}. 
To simulate the generation of CME and measure the final observable, we consider the non-zero chiral chemical proportional to $\mu_5$ using Eq.\eqref{mu5eq}. Since the current of CME is proportion to $\mu_5$, the results with zero $\mu_5$ can be thought as background which contains all effects except the CME.


\section{Parameters setting for isobar collisions}
\label{isobarcollision}
In order to overcome the large uncertain backgrounds, we study the isobar $_{44}^{96}\text{Ru}+\rm{} _{44}^{96}Ru$ and $_{40}^{96}\text{Zr}+\rm{}_{40}^{96}Zr$ with CAT module. There are similar backgrounds in isobar collisions because of the same mass number of these two nuclei, while the more proton in Ru should induce larger magnetic field and then induce larger CME signal in Ru+Ru collisions \cite{Voloshin:2010ut,Deng:2016knn}.
 It is expected that the different two-particle correlations $\Delta \gamma$ will be observed within these collision systems.

But as the isobar blind analysis in the STAR collaboration \cite{STAR:2021mii} progresses, the backgrounds are observed not identical. The elliptic flow $v_2$ and multiplicity $N_{ch}$ are different between two collisions systems. The results of this research show the $\Delta\gamma$ unexpectedly smaller in Ru+Ru than in Zr+Zr collisions and the differences of backgrounds in isobar collisions become more significant. The most notable disparities between Ru+Ru and Zr+Zr collisions stem from the number of protons and the nuclear structure. The variation in the number of protons affects the magnetic field, which in turn influences the evolution of the Quark-Gluon Plasma (QGP). The nuclear structure directly impacts the initial geometric anisotropy of the transverse overlap region and the distribution of spectator protons. Both of these factors contribute to altering the final-state observables, such as flow and multiplicity.

\subsection{Nuclear structure}

The spatial distribution of nucleons in the rest frame of  $_{44}^{96}$Ru and $_{40}^{96}$Zr plays a key role to explain the differences in two collisions systems. Extensive previous research \cite{Li:2024aeu,STAR:2024wgy,Jia:2022qgl} has been dedicated to understanding the impact of nuclear structure on the backgrounds of the Chiral Magnetic Effect (CME). These studies indicate that the ratios of elliptic flow and charge multiplicity between Ru+Ru and Zr+Zr are primarily influenced by their nuclear structures. However, the precise nuclear structures, particularly their deformations, remain uncertain at present.

In AMPT model, this distribution can be described by the following Woods-Saxon (WS) form :
\begin{equation}
    \rho(r,\theta)=\frac{\rho_{0}}{1+\exp[(r-R_{0}[1+\beta_{2}Y_2^0(\theta)])/a]},
\end{equation}
where $r$ is the radial position, $\rho_{0} = 0.16\,\mathrm{fm}^{-3}$ is the normal nuclear density, $R_{0}$ is the radius of the nucleus, $\beta_2$ is the deformity quadrupole parameter, $Y_2^0(\theta)$ is the second-order spherical harmonic function and $a$ is the surface diffuseness parameter. As shown in Table-\ref{table of WS parameters}, we use four sets of WS parameters. The case-1, in which the nuclear structures of Ru and Zr are identical, is used to compare the effect of different the magnetic field. The case-2 is based on the calculations of the energy density functional theory \cite{Xu:2017zcn} to investigate the background effect from nuclear structure. 
In case 3, we employ a non-zero $\beta_2$ to examine quadrupole deformation and its impact on the magnetic field. However, these deformation parameters are unsuitable when analyzing the ratio of  $N_{ch}$ and $v_2$ in Ru+Ru collisions compared to Zr+Zr collisions.
In the case-4, the nuclear structure of proton and neutron in Zr is different, which is given by the neutron skin effect coming from the differences of density distribution between proton and neutron \cite{Xu:2021vpn,Xu:2017zcn}.
It should be noted that the parameters for Ru in cases 1, 2, and 4 are identical, resulting in the same observables. Therefore, we will use the same markers or a single line to represent the results of these cases.
To determine which cases effectively describe the geometric differences between the two nuclei, we will focus on the charged particle multiplicity ($N_{ch}$) and the elliptic flow ($v_2$), comparing their ratio with STAR data. However, it is essential to first consider the differences in the magnetic field.

\begin{table}[h!]
  \begin{center}
    \caption{table of WS parameters\cite{Xu:2021vpn,Xu:2017zcn}}
      \label{table of WS parameters}

    \begin{tabular}{l|r|c|c|c}
      \textbf{case} & \textbf{Nucleus} & \textbf{R(fm)} & \textbf{a(fm)} & $\bf{\beta_2} $ \\
      \hline
      
      \multirow{2}{*}{1}& Ru  & 5.085& 0.523 &0\\
      & Zr & 5.085& 0.523 &0\\
            \hline
       \multirow{2}{*}{2}& Ru  & 5.085& 0.523 &0\\
      & Zr & 5.021& 0.592 &0\\
                  \hline
      \multirow{2}{*}{3}& Ru  & 5.085& 0.523 &0.2\\
      & Zr & 5.021& 0.592 &0.2\\
                  \hline
      \multirow{3}{*}{4}& Ru  & 5.085& 0.523 &0\\
      & Zr n & 5.021& 0.592 &0\\
      & Zr p & 5.021& 0.523 &0\\
    \end{tabular}
  \end{center}

\end{table}

 
 


\subsection{Magnetic field}
\begin{figure}[!ht]
\hspace*{-6 mm}
\includegraphics[scale=0.38]{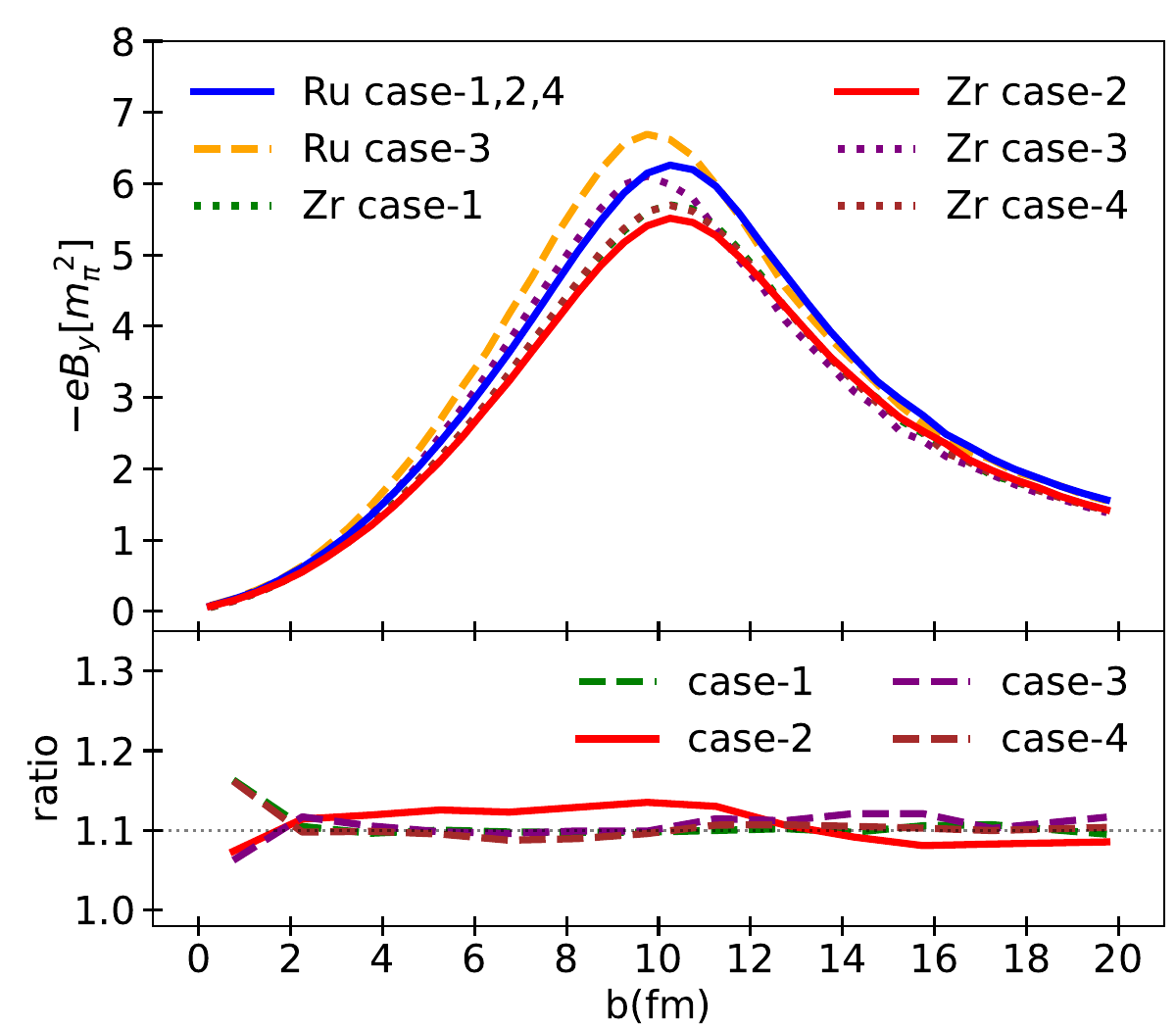}
\caption{CAT results for the initial magnetic field along the y-axis ($eB_y$) at the center point of the fireball are shown in the upper panel. The lower panel displays the ratio of $eB_y$ in Ru+Ru collisions to that in Zr+Zr collisions as a function of the impact parameter b. These results are presented for four different Woods-Saxon parameter settings in 200 GeV isobar collisions.}
\label{ebsiobar}
\end{figure}

\begin{figure}[!ht]
\hspace*{-6 mm}
\includegraphics[scale=0.40]{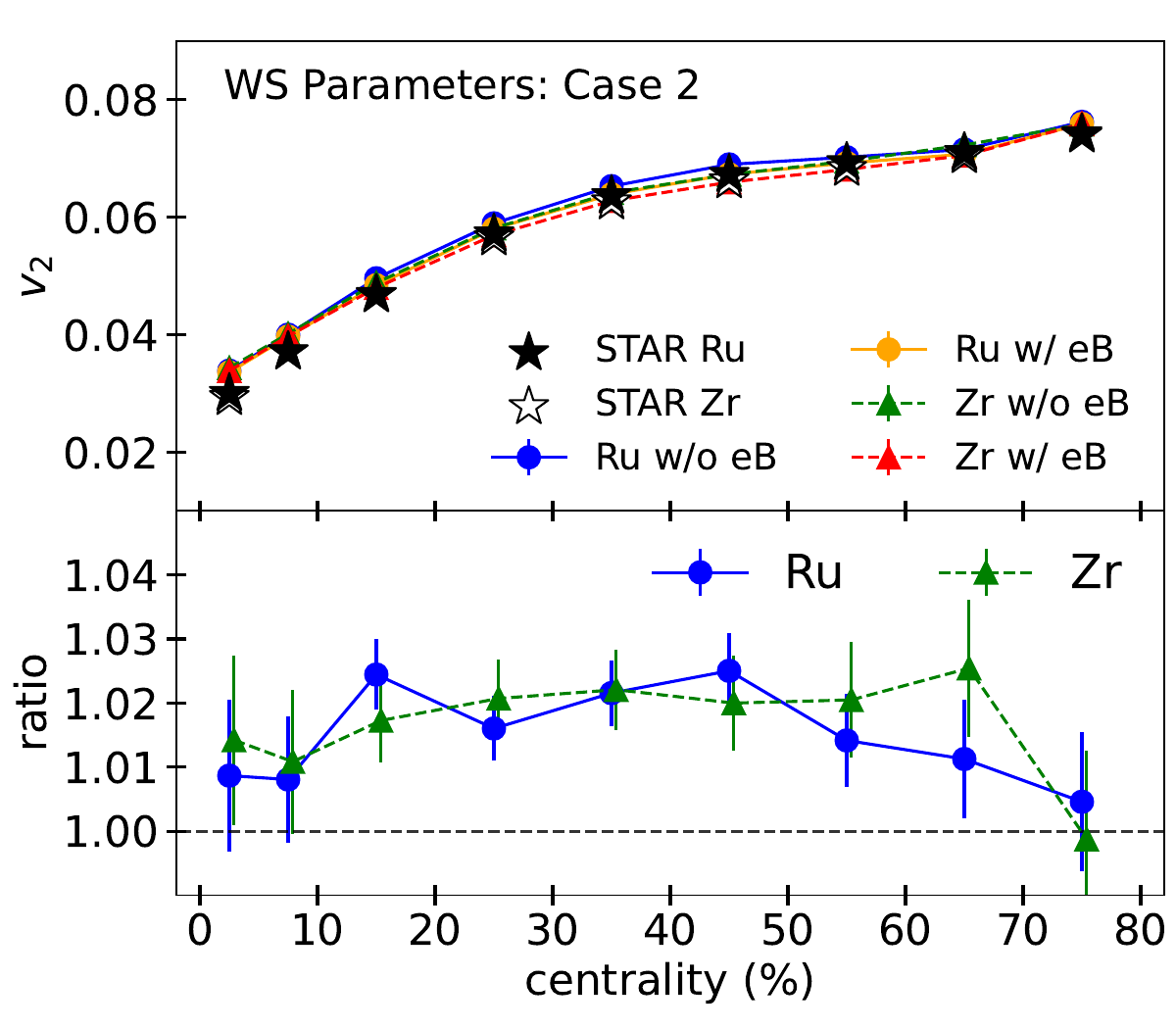}
\caption{
(Upper panel) The results from the CAT, illustrating the centrality dependence of $v_{2}\{EP\}$ for both cases: with a magnetic field (solid lines) and without a magnetic field (dashed lines). These results are compared with experimental data from STAR for Ru+Ru and Zr+Zr collisions. (Lower panel) The ratio of $v_{2}\{EP\}$ obtained without a magnetic field to that with a magnetic field is shown.}
\label{isoeb2}
\end{figure}

The upper panel of Fig.\ref{ebsiobar} illustrates the magnitude of the initial magnetic field $eB_0$ at the center of the two collision systems. This is calculated using CAT with Eq.\eqref{mg12} for the four different parameter settings. The differences of the magnetic field between two collision systems are evident in all cases \cite{Zhao:2019crj}. 
The lower panel of Fig.\ref{ebsiobar} shows the ratio of magnetic field  $eB_{y,Ru}\approx 1.1 eB_{y,Zr}$ for all cases of isobar collisions, while the ratio for case-2 is slight larger when $b<12$ fm and then becomes little smaller. 
The magnetic filed of Zr is almost the same between case-1 and case-4,due to the same parameters of proton. And the smaller effective radius in case-2 induced by larger parameter $a$ leads to little smaller magnetic filed shown in Fig.\ref{ebsiobar}, when $b<12$ fm. Comparing the results of case-1 and case-3, the non-zero deformation $\beta_2$ makes magnetic field increase. These results suggest that the number of protons primarily determines the magnitude of the initial magnetic field and the nuclear structure, particularly the quadrupole deformation, still introduces variations in the magnetic field. These variations are then transmitted to the background and CME signal.

Our previous research on Au+Au collisions demonstrated that the presence of a magnetic field reduces the elliptic flow $v_2$. Consequently, we have investigated the impact of the magnetic field on $v_2$ in isobar collisions. The upper panel of Figure \ref{isoeb2} illustrates that when the magnetic field is zero, the $v_2$ for Ru and Zr in case-2 is slightly higher compared to when the magnetic field is considered. The lower panel of Figure \ref{isoeb2} show the ratios of $v_2$ when magnetic field is zero to that with a non-zero magnetic field and  reveals that the magnetic field results in a $1\sim3\%$ difference in Ru+Ru and Zr+Zr collisions. The discrepancy in $v_2$ caused by the magnetic field is expected to be proportional to the strength of the magnetic field and $eB_{Ru+Ru}\approx 1.1 eB_{Zr+Zr}$. Therefore, the varying magnetic field could reduce the ratio of $v_2$ in Ru+Ru collisions to that in Zr+Zr collisions, and slightly diminish the influence of the initial geometric anisotropy on the $v_2$ ratio, while this discrepancy is weak compared with statistic error. Regarding the charged particle multiplicity, the magnetic field has a negligible effect.

\subsection{Elliptic flow and mean charged multiplicity}

\begin{figure}[!ht]
\hspace*{-6 mm}
\includegraphics[scale=0.53]{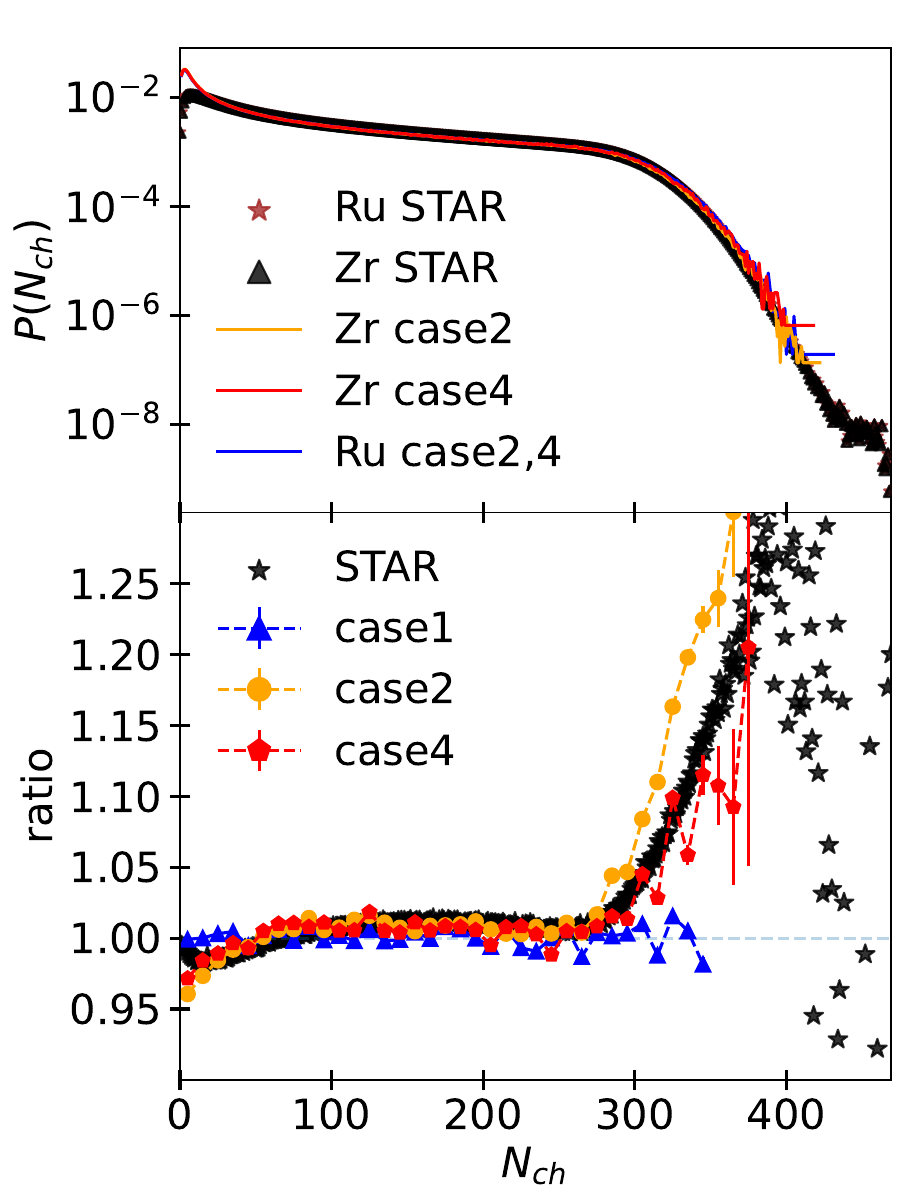}
\caption{(Upper panel) Distributions of the number of charged particles ($N_{ch}$) from CAT module (lines) for case-2 and case-4 in the pseudorapidity acceptance $|\eta|<0.5$ in Ru+Ru and Zr+Zr collisions compared with STAR data (dots). These distributions are corrected for normalization \cite{Nijs:2021kvn}.
(Lower panel) The Ru+Ru to Zr+Zr $N_{ch}$ ratios of model for case-2 and case-4 compared with STAR data (stars). }
\label{ndistribution}
\end{figure}

Fig.\ref{ndistribution} shows the distributions of $N_{ch}$ (upper panel) and ratios of $N_{ch}$ (lower panel) in Ru+Ru and Zr+Zr collisions from CAT simulation for case-2 and case-4 compared with STAR data. These distributions of $N_{ch}$ are almost identical with the STAR multiplicity after multiplying by a normalization correction \cite{Nijs:2021kvn}. This correction arises from the difference between the low efficiency of STAR low multiplicity events detection and high efficiency of simulation. This correction is defined as the ratio of the integral for $N_{ch}>10$ distribution from CAT to that from STAR data. The distributions for case-1 and case-3 are also similar to STAR data, which are not drawn here for brevity. Shown in the lower panel of Fig.\ref{ndistribution}, the ratio of $N_{ch}$ in Ru+Ru collisions to that in Zr+Zr collisions is nearly one for case-1 in which the nuclear structure parameters of Ru and Zr are the same and induce the same multiplicity distribution. The ratio of $N_{ch}$ for case-2 can describe STAR data in the $15<N_{ch}<280$ range. For case-4, the ratio is similar to STAR data in the $15<N_{ch}<350$ range. The results of case-3 is similar to the results of case-4. Because the $N_{ch}<20$ range  represents the peripheral collisions in $70-100\%$ centrality range and $N_{ch}>300$ range represents the most central collisions in $0-5\%$ centrality range, these results of ratio are suitable if we only consider the middle centrality, 

Because the centrality interval is defined as the $N_{ch}$ percentile of the total event distribution, the mean charged multiplicity $\left \langle N_{ch}  \right \rangle $ for a certain centrality range directly depends on the centrality definition. The slightly deviation from 1 of the ratios of $N_{ch}$ in Fig.\ref{ndistribution} for case-2,4 induce the different centrality definition in the two collision systems. Thus,  $\left \langle N_{ch}  \right \rangle_{Ru}$ is not identical to $\left \langle N_{ch}  \right \rangle_{Zr}$ for case-2, 3 and 4. 

\begin{figure}[!ht]
\hspace*{-6 mm}
\includegraphics[scale=0.40]{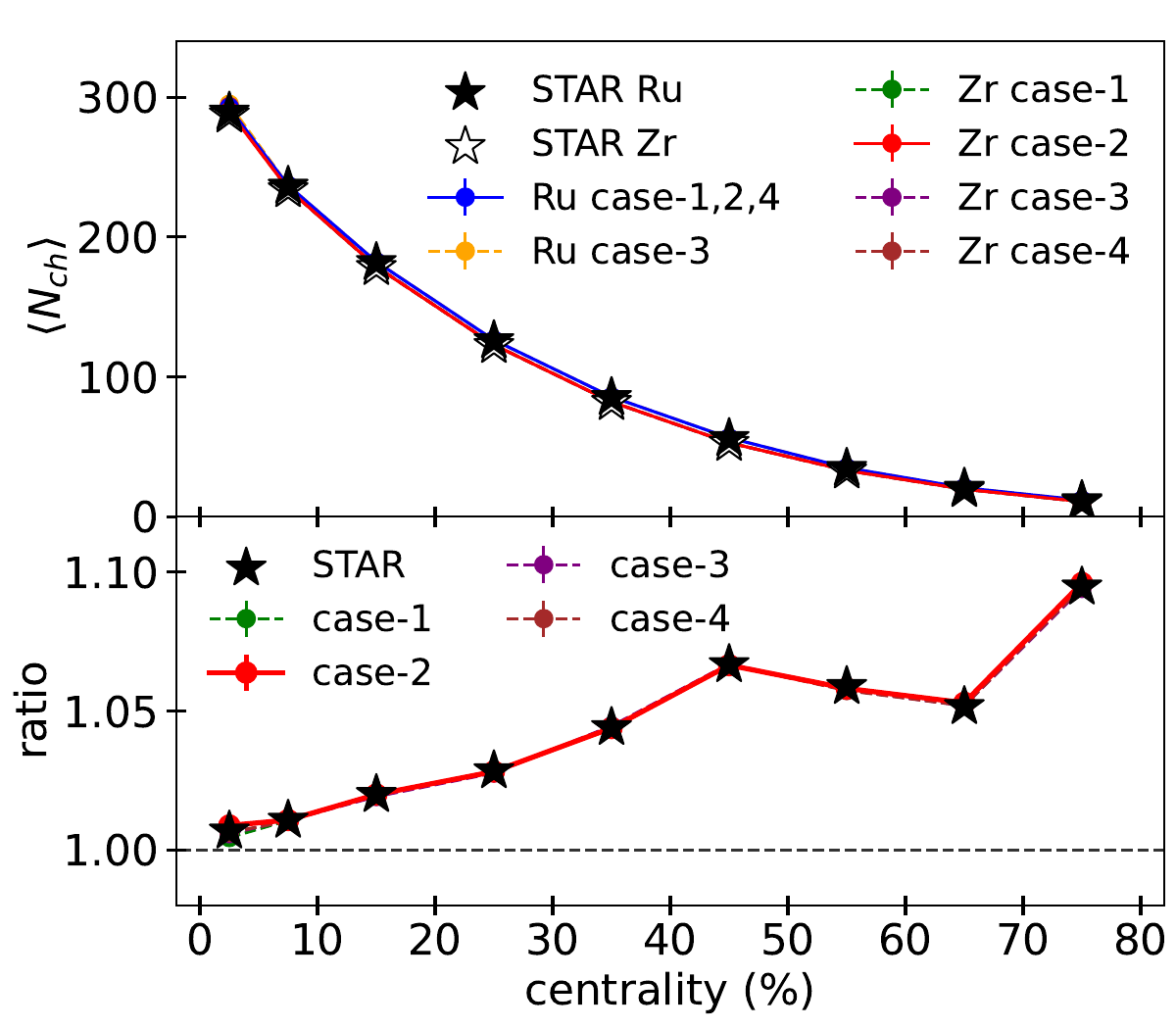}
\caption{(Upper panel) The mean number of charged particles $\left \langle N_{ch}  \right \rangle $  as a function of centrality in Ru+Ru and Zr+Zr collisions from CAT (lines) for four cases compared with STAR data (stars). (Lower panel) The centrality dependence of the ratio of $\left \langle N_{ch}  \right \rangle $  in Ru+Ru collisions to that in Zr+Zr collisions from CAT (lines) for four cases compared with STAR data (stars). The centrality definitions are identical to STAR experiment for all cases. The statistical uncertainties are so small that are ignored. }
\label{nch star}
\end{figure}

In this paper, we use the original multiplicity distributions from CAT to define the centrality for different cases and different collision systems, but STAR has used the experimental data and the results of MC Glauber simulations for $0-20\%$ centrality and $20-100\%$ centrality\cite{STAR:2021mii}, separately. Although the ratios of CAT are similar to STAR, the definitions of centrality for the same nuclei between CAT simulation and STAR measurement are subtly different, due to the slight differences.
If the centrality definitions in this simulation were the same as STAR, the mean charged multiplicity $\left \langle N_{ch}  \right \rangle $ and the ratio of  $\left \langle N_{ch}  \right \rangle $ in Ru+Ru collisions to that in Zr+Zr collisions would be identical to STAR data for all cases, as shown in Fig.\ref{nch star}. We expect the ratio of $N_{ch}$ could reflect the effect of nuclear structure. But the same values of $N_{ch,Ru}/N_{ch,Zr}$ in the lower panel of Fig.\ref{nch star} for STAR data and simulation results generated from the same centrality interval obscure this effects and the ratio of $N_{ch}$ for case-1 incorrectly deviated by one. So, in this paper, we use centrality defined by CAT results instead of the centrality defined by STAR.

\begin{figure}[!ht]
\hspace*{-6 mm}
\includegraphics[scale=0.40]{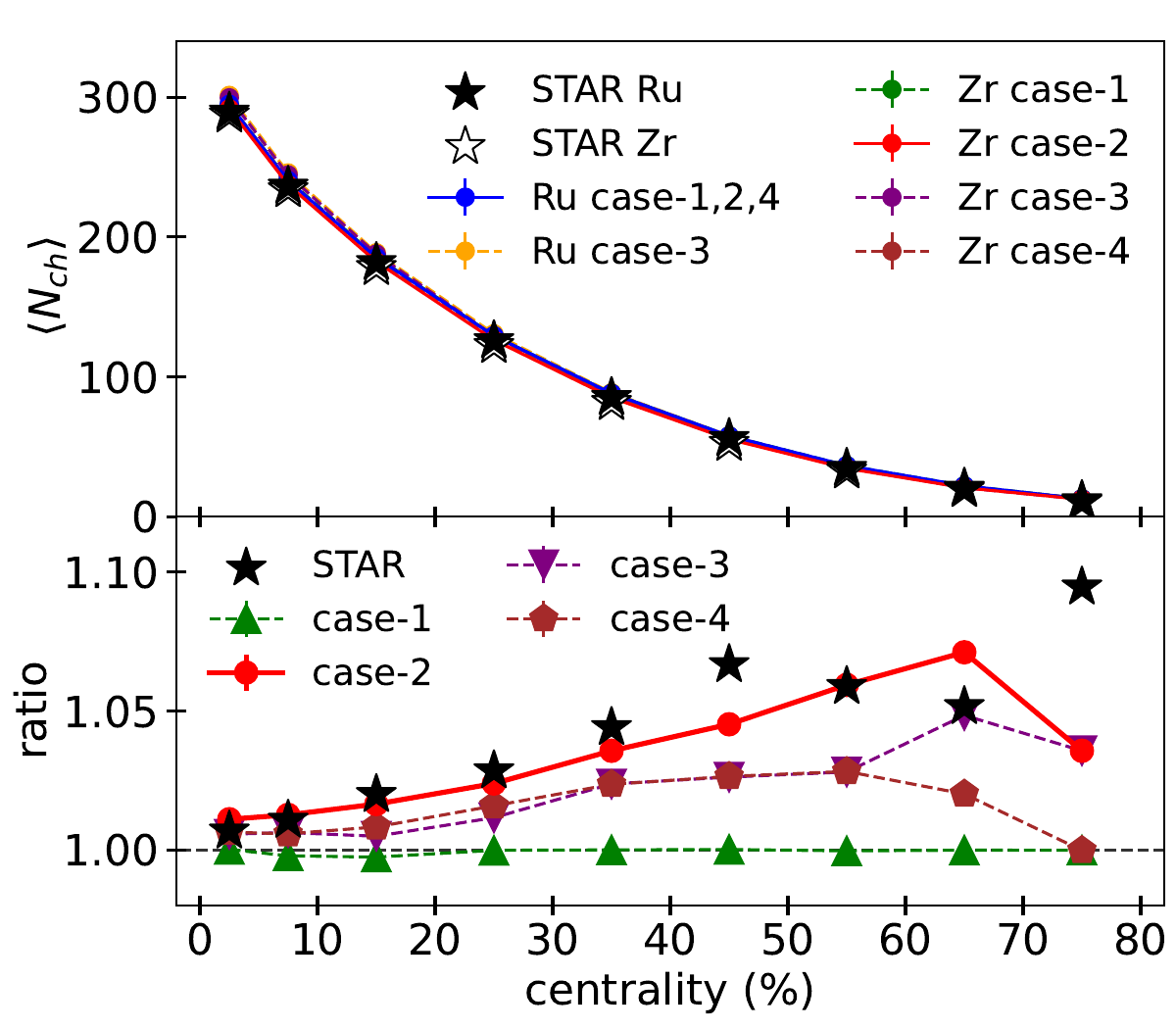}
\caption{(Upper panel) The mean number of charged particles $\left \langle N_{ch}  \right \rangle $  as a function of centrality in Ru+Ru and Zr+Zr collisions from CAT (lines) for four cases compared with STAR data (stars). (Lower panel) The centrality dependence of the ratio of $\left \langle N_{ch}  \right \rangle $  in Ru+Ru collisions to that in Zr+Zr collisions from CAT (lines) for four cases compared with STAR data (stars).The centrality is defined by the CAT multiplicity. }
\label{nchisobar}
\end{figure}

The upper panel of Fig.\ref{nch star} and Fig.\ref{nchisobar} show the mean charged multiplicity $\left \langle N_{ch}  \right \rangle $  within the pseudorapidity window of $|\eta|<0.5$ as a function of centrality from CAT for four cases, compared with STAR data. The simulation values for the four cases of the two nuclei align well with the STAR data. The lower panel of Fig.\ref{nchisobar} shows the centrality dependence of the ratio of the mean charged
 multiplicities in Ru+Ru collisions to that in Zr+Zr col
lisions. This indicates that the mean charged multiplicity ratio for case-1 is approximately unity across the $0-80\%$ centrality range, consistent with the charged multiplicity distribution. This suggests that the number of protons does not influence the average charged multiplicity, $\left \langle N_{ch}  \right \rangle$.
The ratio for case-2 better conforms to the STAR data than other cases, but it is smaller than data in the $40-50\%$ and $70-80\%$ centrality. The ratio for case-3 and case-4 are smaller than STAR data. These differences between simulation results and STAR data can be attributed to the inappropriate nuclear structure and the difference in the centrality definitions between our work and the STAR data, which we have previously discussed. If the same centrality definition is used, the values of $\left \langle N_{ch}  \right \rangle $ and their ratios align with the STAR data. 

\begin{figure}[!ht]
\hspace*{-6 mm}
\includegraphics[scale=0.4]{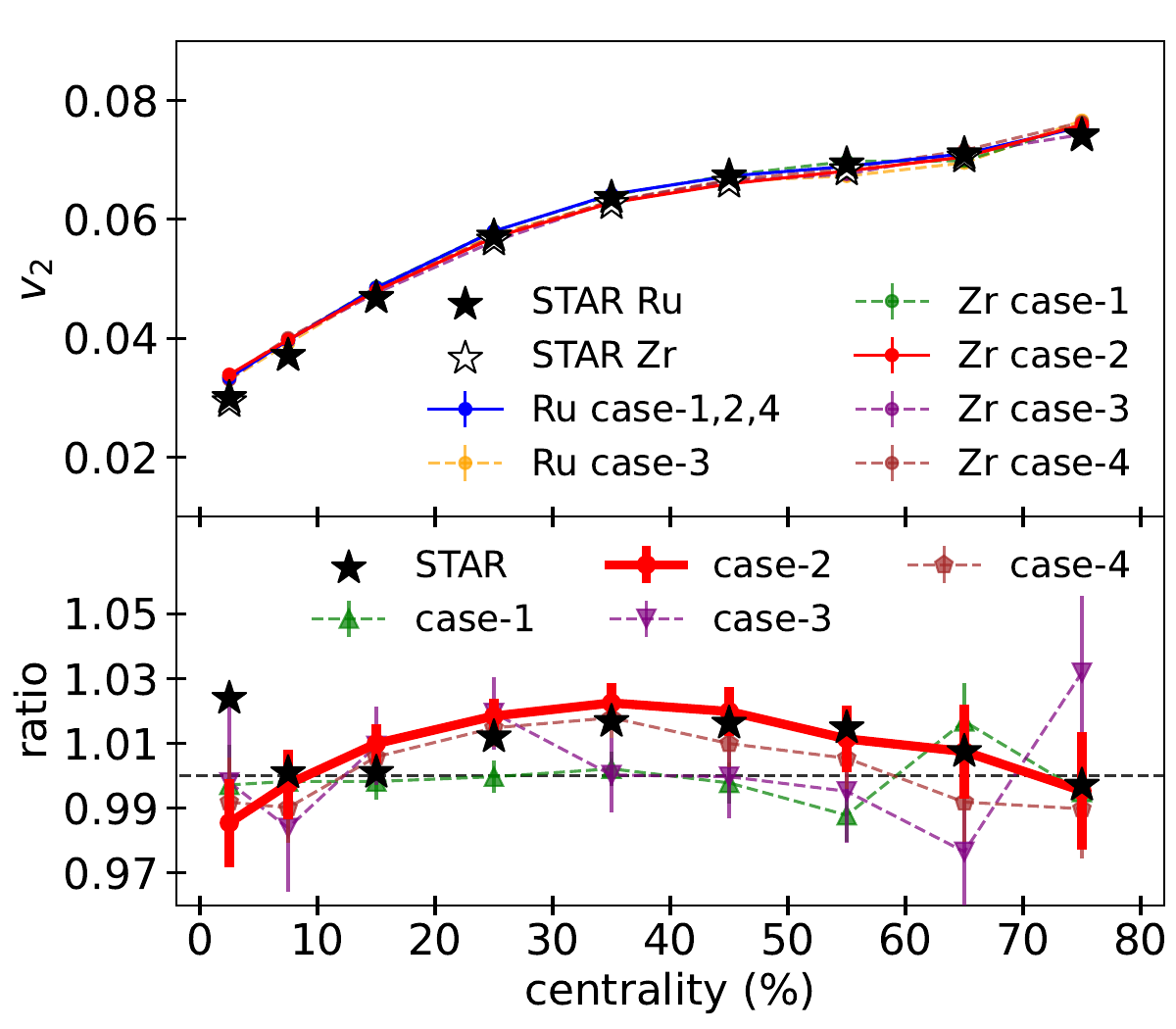}
\caption{(Upper panel) The elliptic flow $v_2$ using sub-event methods as a function of centrality in Ru+Ru and Zr+Zr collisions from CAT (lines) for four cases compared with STAR data (stars). (Lower panel) The centrality dependence of the ratio of $v_2$ in Ru+Ru collisions to that in Zr+Zr collisions from CAT (lines) for four cases compared with STAR data (stars). }
\label{v2cases}
\end{figure}

Shown in Fig.\ref{v2cases}, the upper panel shows the different centrality dependence of the elliptic flow $v_2$ of particles in the $0.2<p_T<2 \ \text{GeV},\,|\eta|<1 $ range in Ru+Ru collisions and Zr+Zr collisions and the ratio of elliptic flow is shown in the lower panel for four cases using event plane method \cite{STAR:2009tro,STAR:2004jwm}. For case-1, the ratio shows nearly one within error in the middle centrality and the influence of different magnetic field are hardly extracted from the statistics uncertainties. For case-2, the simulation results accurately describe the ratio of $v_2$ within the $10-80\%$ centrality range, except for the most central collisions. For case-3, it shows that the quadrupole deformation depresses the ratio. For case-4, the ratio is smaller than that for case-2 and near STAR data in middle centrality, but becomes more smaller in peripheral collisions. Considering the background of CME is related to $N_{ch}$ and $v_2$ and the better result of case-2, we use the parameters of case-2 in the search of CME signal.

\section{CME signal in isobar collisions}
\subsection{CME signal with $\Delta\gamma$}
\begin{figure}[!ht]
\hspace*{-6 mm}
\includegraphics[scale=0.4]{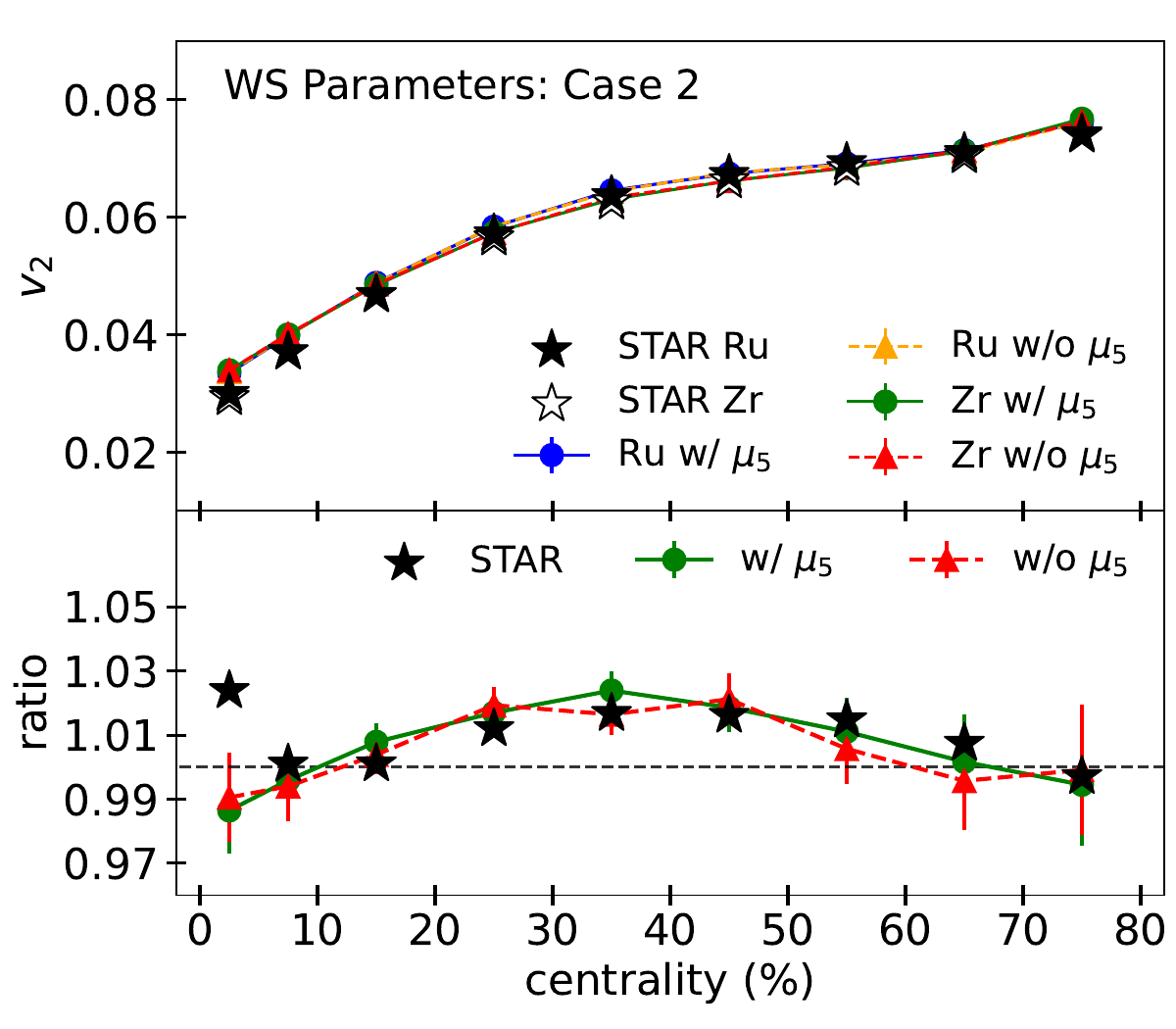}
\caption{(Upper panel) The elliptic flow $v_2$ as a function of centrality in Ru+Ru and Zr+Zr collisions from CAT (lines) with or without $\mu_5$ for cases-2 compared with STAR data (stars). (Lower panel) The centrality dependence of the ratio of $v_2$ from CAT (lines) with or without $\mu_5$ for cases-2 compared with STAR data (stars). }
\label{isov2mu5}
\end{figure}

\begin{figure}[!ht]
\hspace*{-6 mm}
\includegraphics[scale=0.4]{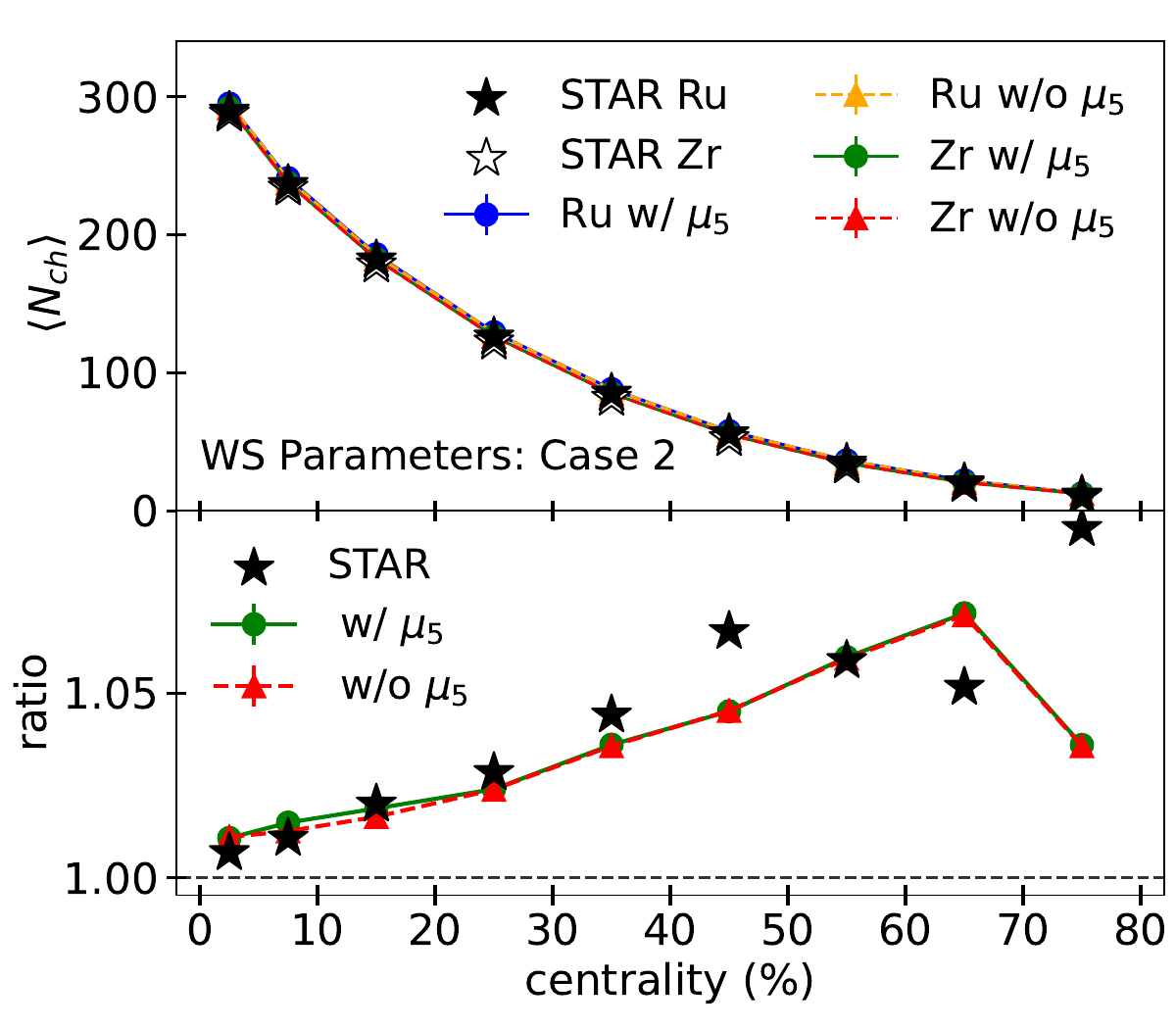}
\caption{(Upper panel) The number of charged particles $N_{ch}$ as a function of centrality in Ru+Ru and Zr+Zr collisions from CAT (lines)with or without $\mu_5$ for cases-2 compared with STAR data (stars). (Lower panel) The centrality dependence of the ratio of $N_{ch}$ from CAT (lines) with or without $\mu_5$ for cases-2 compared with STAR data (stars).  }
\label{nchmu5}
\end{figure}

The background signals in isobar collisions are sensitive to initial conditions. Therefore, it is essential to examine the effect of the chiral chemical potential on these backgrounds. To achieve this, we compared scenarios with and without the initial chiral chemical potential, $\mu_{5}$, as described by Eq.\eqref{mu5eq}. Fig.\ref{isov2mu5} and Fig.\ref{nchmu5} present the centrality dependence of $v_{2}$, $N_{ch}$, and their ratio, both with and without the inclusion of $\mu_5$. The slight variations observed, which fall within the error margins, suggest that the chiral chemical potential's influence on $v_{2}$ and $N_{ch}$ is negligible. This finding allows us to explore the impact of $\mu_{5}$ on the CME without altering the background consistency.

\begin{figure}[!ht]
\hspace*{-6 mm}
\includegraphics[scale=0.4]{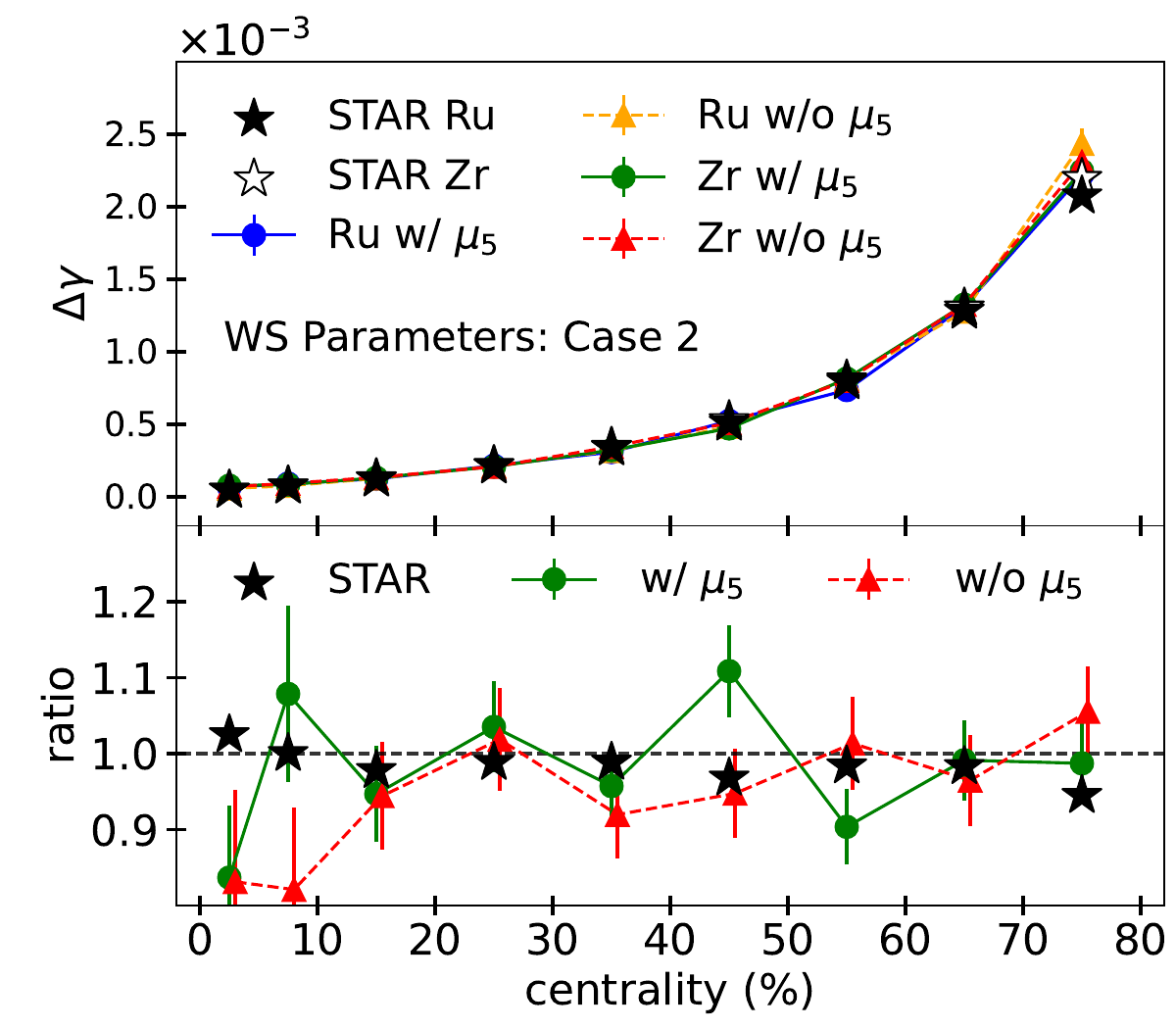}
\caption{(Upper panel) The correlation $\Delta\gamma$ as a function of centrality in Ru+Ru and Zr+Zr collisions from CAT (lines) with or without $\mu_5$ for cases-2 compared with STAR data (stars). (Lower panel) The centrality dependence of the ratio of $\Delta\gamma$ in Ru+Ru collisions to that in Zr+Zr collisions from CAT (lines) with or without $\mu_5$ for cases-2 compared with STAR data (stars).}
\label{cmemu5cen3}
\end{figure}

\begin{figure}[!ht]
\hspace*{-6 mm}
\includegraphics[scale=0.4]{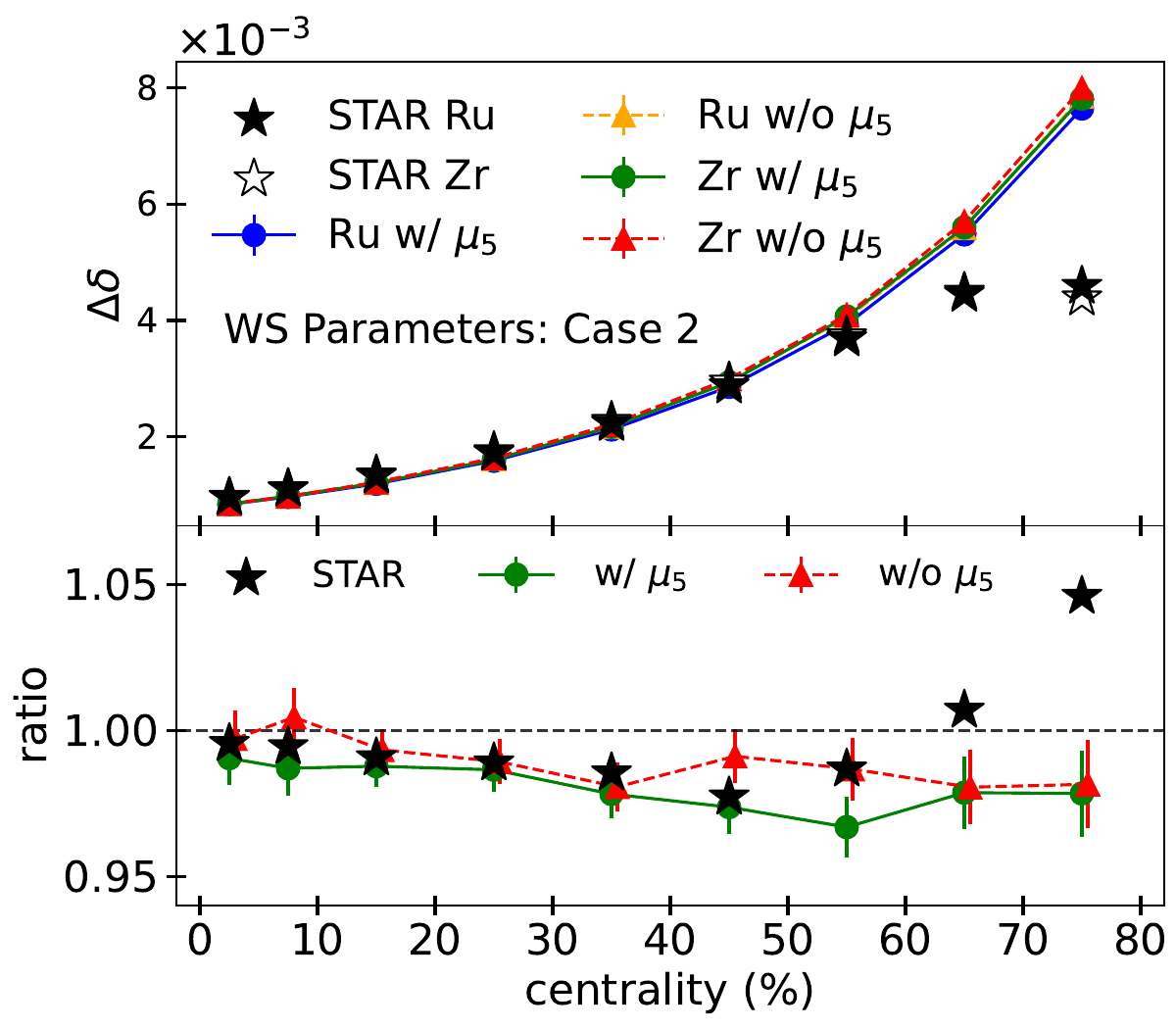}
\caption{(Upper panel) The correlation $\Delta\delta$ as a function of centrality in Ru+Ru and Zr+Zr collisions from CAT (lines) with or without $\mu_5$ for cases-2 compared with STAR data (stars) taken from \cite{STAR:2021mii}. (Lower panel) The centrality dependence of the ratio of $\Delta\delta$ in Ru+Ru collisions to that in Zr+Zr collisions from CAT (lines) with or without $\mu_5$ for cases-2 compared with STAR data (stars).}
\label{isodelta}
\end{figure}


To identify the background, we consider $\Delta\gamma$ with or without $\mu_{5}$ as the total observable $\Delta\gamma_{total}$ or the pure background $\Delta\gamma_{bkg}$, respectively. According to the theoretical prediction, if the fraction of CME is larger in Ru than that in Zr, then the ratios of $\Delta\gamma_{total}$ between two nuclei should larger than the ratios of $\Delta\gamma_{bkg}$. The centrality dependence of $\Delta\gamma$ for the two isobar collisions from the CAT module with or without $\mu_5$ compared with the experimental data are shown in upper panel of Fig.\ref{cmemu5cen3}, where we generate about 20 million events for each case.
The values of $\Delta \gamma$ for the different cases are comparable and align with the STAR data, except in the $70-80\%$ centrality range. Here, the larger $\Delta\gamma_{total}$ and $\Delta\gamma_{bkg}$ compared to the STAR data may be caused by a slightly higher background of $v_2$, as shown in Fig.\ref{v2cases}, and differences in centrality definitions between our work and STAR. The lower panel of Fig.\ref{cmemu5cen3} shows the centrality dependence of the ratio of $\Delta\gamma$ with or without $\mu_{5}$. The ratio $\Delta\gamma_{Ru,bkg}/\Delta\gamma_{Zr,bkg}$ shows a tendency of less than one within large errors. The ratio $\Delta\gamma_{Ru,total}/\Delta\gamma_{Zr,total}$ tends to be larger than $\Delta\gamma_{Ru,bkg}/\Delta\gamma_{Zr,bkg}$ as expected, except in centrality $20-30\%$ and $70-80\%$, where the values of this ratio are acceptable within the error range.

As mentioned above, the reaction plane-independent correlation \(\Delta\delta\) is crucial for interpreting background correlations. We present the reaction plane-independent correlation \(\Delta\delta\) (upper panel) and their ratios (lower panel), compared with STAR data from \cite{STAR:2021mii}, as shown in Fig.\ref{isodelta}. The simulation's result for $\Delta\delta$ aligns well with the STAR data across the mid-centrality range, except for the $60-80\%$ centrality. For the ratio of $\Delta\delta$ between Ru+Ru collisions and Zr+Zr collisions, the simulation results, both with and without $\mu_5$, are in close agreement with the STAR data within the $0-50\%$ centrality range. However, they do not capture the trend observed in the data for the $60-80\%$ centrality range. These discrepancies imply that the model simulation overlooked some effect in the  peripheral centrality. Unlike $\Delta\gamma$, $\Delta\delta$ is not related to reaction plane, so that the resolution is enough to study the effect of $\mu_5$.
Seen in the lower panel of Fig.\ref{isodelta}, the non-zero chiral chemical potential $\mu_5$ reduces the ratio of $\Delta\delta$. And noticed that the CME should reduce the  $\Delta\delta$, there are larger CME signal in Ru+Ru collisions in the simulation as we expect. However, the signal appears much smaller than expected.

 
 \begin{figure}[!t]
\centering
\subfloat{
		\includegraphics[scale=0.4]{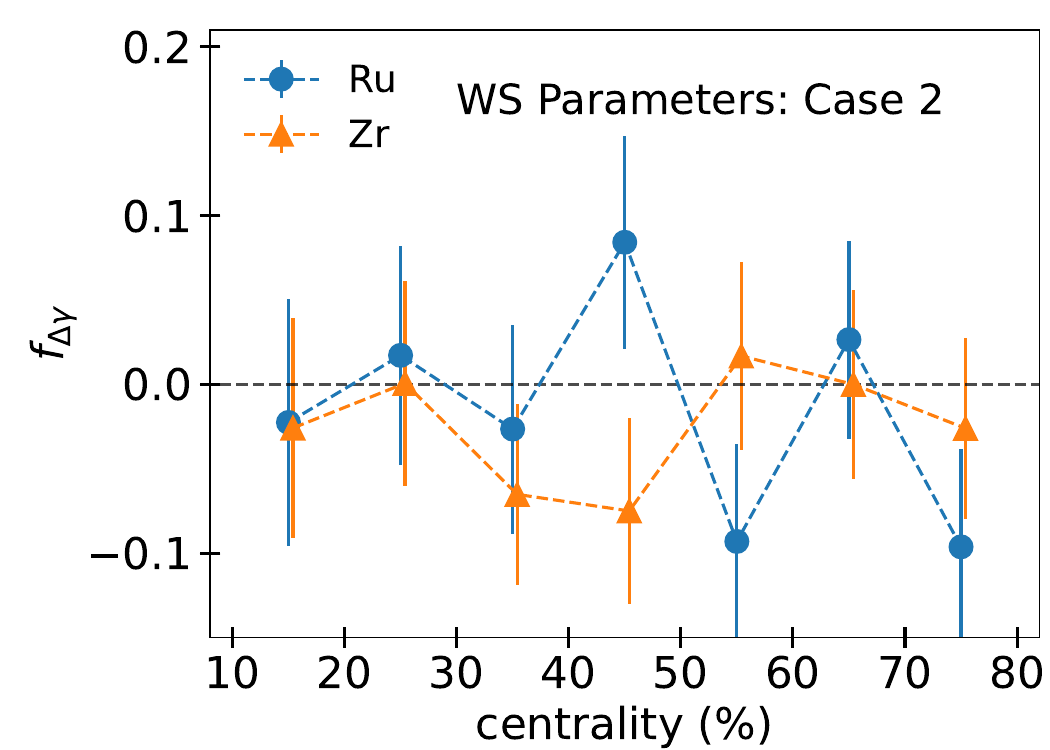}}\\
\subfloat{
		\includegraphics[scale=0.4]{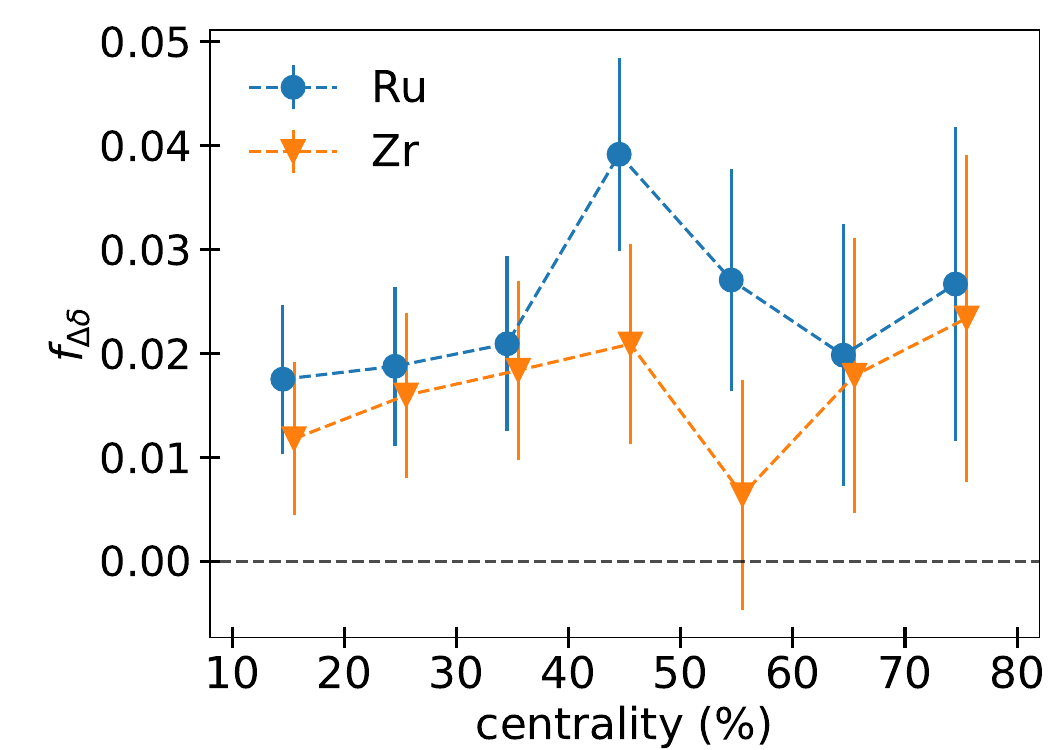}}
\caption{The CAT result on the centrality dependence of the CME fraction $ f_{CME} $ in Ru+Ru collisions and Zr+Zr collisions.}
\label{cmefractioncen3}
\end{figure}

Fig.\ref{cmefractioncen3} shows the CME fraction in isobar collision derived from Fig.\ref{cmemu5cen3} using the following equation:
 \begin{equation}\label{fcme}
 \begin{split}
f_{\Delta\gamma}=\frac{\Delta\gamma_{CME}}{\Delta\gamma_{total}}
=\frac{\Delta\gamma_{total}-\Delta\gamma_{bkg}}{\Delta\gamma_{total}}\\
f_{\Delta\delta}=\frac{\Delta\delta_{CME}}{\Delta\delta_{total}}
=\frac{\Delta\delta_{bkg}-\Delta\delta_{total}}{\Delta\delta_{total}}
 \end{split}
\end{equation}
 
 Unlike Au+Au collisions, the CME fractions in isobar collisions are considerably smaller and obscured by significant errors. This is due to the weaker magnetic field, smaller collision system, and larger statistical fluctuations. Taking these errors into account, the small CME fractions range from -10$\%$ to 15$\%$.
On the other hand, the minor differences observed with or without $\mu_5$
 indicate that the CME observable $\Delta\gamma$ is not sufficiently sensitive to the chiral anomaly to detect the small CME signal from the large background. This implies that we may need to explore other observables. To extract a more precise fraction using $\Delta \gamma$, future CAT simulations will need to generate $1\times10^{9}$ events, which will require significant time and resources similar to the original AMPT\cite{Zhao:2022grq,Chen:2023jhx}.
 

 \subsection{CME signal with $R_{\Psi_{2}}(\Delta S)$}
 A newly developed correlation, $R_{\Psi_{2}}(\Delta S)$, serves as a promising new observable for the CME, expressed as a ratio\cite{Magdy:2017yje}:
\begin{equation}
    R_{\Psi_{2}}(\Delta S)=C_{\Psi_{2}}(\Delta S)/C^\perp _{\Psi_{2}}(\Delta S),
\end{equation}
  where correlation function $C_{\Psi_{2}}(\Delta S)$ and $C^\perp _{\Psi_{2}}(\Delta S)$ are constructed to quantify charge separation $\Delta S$, parallel and perpendicular (respectively) to the $\Psi_{2}$ which is approximately perpendicular to magnetic field. Because CME induces charge separation $\Delta S$ along the magnetic field, this separation only contributes to $C_{\Psi_{2}}(\Delta S)$, and other separations driven by the background effect contribute to both $C_{\Psi_{2}}(\Delta S)$ and $C^\perp _{\Psi_{2}}(\Delta S)$. These correlation functions are constructed as the ratio:
\begin{equation}
    C_{\Psi_{2}}(\Delta S)=N_{real}(\Delta S)/N_{shuffled}(\Delta S),
\end{equation}
where $N_{real}(\Delta S)$ is the distribution of the event-by-event charge separation $\Delta S$:
\begin{equation}
   \Delta S=\frac{\sum_{1}^{p}\omega_i^p \sin(\Delta \phi_i)}{\sum_{1}^{p}\omega_i^p }-\frac{\sum_{1}^{n}\omega_i^p \sin(\Delta \phi_i)}{\sum_{1}^{n}\omega_i^n },
\end{equation}
where $p$ and $n$ are the numbers of positive and negative charge hadrons in an event, $\Delta \phi_i=\phi_i-\Psi_{\text{RP}}$ is the azimuthal emission angle of the charged hadrons. The $N_{shuffled}(\Delta S)$ is the distribution of $\Delta S$ of one event but after random reassignment (shuffling) of the charges of each particle. This shuffle process makes the properties of the numerator and denominator identical except the charge dependence.

 The correlation $R_{\Psi_{2}}(\Delta S)$ measures the correlation of charge separation parallel to the magnetic field, relative to that perpendicular to the magnetic field. In expectation, the charge separation along the the magnetic field leads to a concave-shaped $R_{\Psi_{2}}(\Delta S)$ distributions and a narrower shape of $R_{\Psi_{2}}(\Delta S)$ indicates more evident charge separation along the magnetic field.
 The width $\sigma_{R_{\Psi_{2}}}$ of $R_{\Psi_{2}}(\Delta S)$ distribution measures the strength of CME, but it is influenced by the particle number fluctuations and event plane resolutions. Scaling $\Delta S$ by the width $\sigma_{\Delta sh}$ of $N_{shuffled}(\Delta S)$ distribution, 
 the correlations of $\Delta S^{'}=\Delta S/\sigma_{\Delta sh}$ will minimize the effect of particle number fluctuations. Similarly, the effect of event plane resolutions can be accounted by using $\Delta S^{''}=\Delta S^{'}/\delta_{res}$, where $\delta_{res}$ is the event plane resolution. 
 If the CME-produced charge separation is observed stronger in the Ru+Ru than that in Zr+Zr, the width $\sigma_{R_{\Psi_{2}}}$ of $R_{\Psi_{2}}(\Delta S^{''})$ distribution should be presented as:
 
 \begin{equation}
     1/\sigma^{Ru+Ru}_{R_{\Psi_{2}}}>1/\sigma^{Zr+Zr}_{R_{\Psi_{2}}}
     \label{widthsrelation}
 \end{equation}
 
\begin{figure}[!ht]
\hspace*{-6 mm}
\includegraphics[scale=0.44]{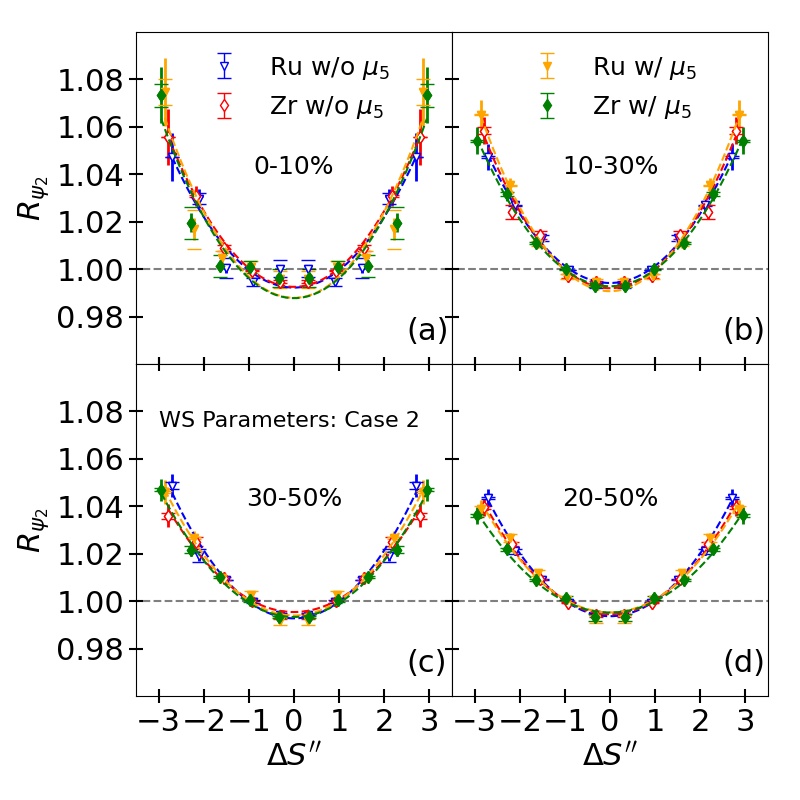}
\caption{Comparison of the $R_{\Psi_{2}}(\Delta S^{''})$ distributions obtained for charged particles in (a) $0-10\%$, (b) $10-30\%$, (c) $30-50\%$ and (d) $20-50\%$ collisions from CAT (dots) and the fits (dashed lines) in Ru+Ru and Zr+Zr collisions at $\sqrt{s_{NN}}$=200 GeV with or without $\mu_5$.The distributions shown in (a)-(d) are symmetrized around $\Delta S^{''}=0$ as same as STAR\cite{STAR:2021mii}.}
\label{rpsi2isobar}
\end{figure}

\begin{figure}[!ht]
\hspace*{-6 mm}
\includegraphics[scale=0.44]{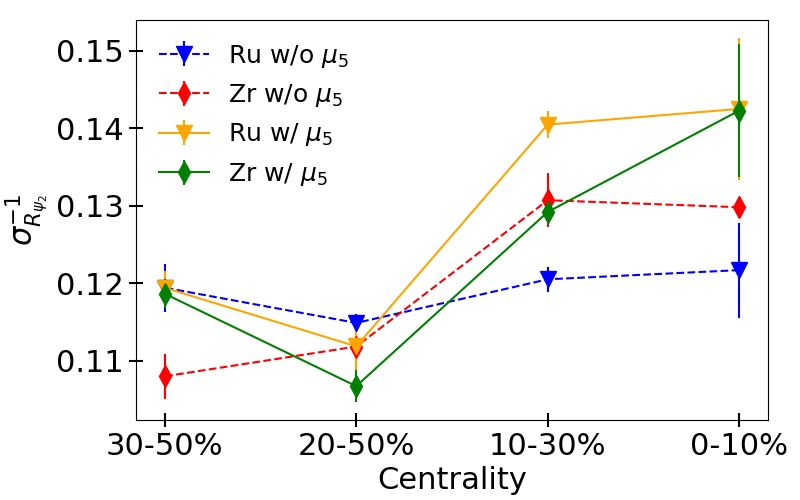}
\caption{The centrality dependence of the inverse widths $\sigma^{-1}_{R_{\Psi_{2}}}$, extracted from the $R_{\Psi_{2}}(\Delta S^{''})$ distributions. }
\label{rpsi2sigma}
\end{figure}
Fig. \ref{rpsi2isobar}(a-d) show the $R_{\Psi_{2}}(\Delta S^{''})$ distributions using sub-event cumulants method and particles within $0.2<p_{T}<2\,\text{GeV},\,|\eta|<1$ in Ru+Ru and Zr+Zr collisions for $0-10\%,10-30\%,30-50\%,20-50\%$ centrality selections, respectively, as same as that in STAR measurement. The dashed lines represent the Gaussian fitting for the simulation data from which we extract the widths $\sigma_{R_{\Psi_{2}}}$. In the most central (a), the different fitting results are induced by the large statistical uncertainties and systematic uncertainties in the $R_{\Psi_{2}}(\Delta S^{''})$ distributions which increase with the magnitude of $\Delta S^{''}$ because the smaller $\sigma_{\Delta sh}$ of $N_{shuffled}(\Delta S)$ distribution magnify the uncertainties compared with that in mid-central. And in the mid-central (d), the $R_{\Psi_{2}}(\Delta S^{''})$ distributions and fit lines are  are nearly identical in the $-2<\Delta S^{''}<2$ range.

The similar shapes of $R_{\Psi_{2}}$ for the two isobars under different conditions suggest that the differences between isobar collisions have only a minor effect.
The centrality dependence of widths $\sigma^{-1}_{R_{\Psi_{2}}}$ for the two isobars are shown in Fig.\ref{rpsi2sigma}. 
As opposed to the STAR results \cite{STAR:2021mii}, the centrality dependences for both isobars decrease as the collisions become more peripheral. But the confidence level of $\sigma^{Ru+Ru}_{R_{\Psi_{2}}}$ to $\sigma^{Zr+Zr}_{R_{\Psi_{2}}}$ is low due to the absence of data uncertainties in the fit measurement.

\section{Conclusion}

In this work, we conducted a detailed investigation of the possible CME signal in isobar collisions of  $_{44}^{96}\text{Ru}+\rm{} _{44}^{96}Ru$ and $_{40}^{96}\text{Zr}+\rm{}_{40}^{96}Zr$, utilizing the newly developed Chiral Anomaly Transport (CAT) module\cite{Yuan:2023skl}which includes the chiral anomaly, the magnetic filed and the generation of CME. Utilizing the CAT module, we conducted an in-depth analysis of how the magnetic field and the nuclear structures of Ru and Zr   influence both the background and the CME signal, with a specific emphasis on the  parameters within their Woods-Saxon (WS) distributions.

The CAT module successfully identifies the impact of nuclear structure and magnetic fields on multiplicity distribution, the average number of charged particles, elliptic flow, and their respective ratios in the mid-centrality range. Utilizing the Woods-Saxon parameters from case-2, the findings provide an excellent description of those experimental quantities. By comparing simulations with and without the Chiral Magnetic Effect (CME), we successfully isolated the CME signal from the background. We observed a negligible influence of the chiral chemical potential on the CME observables $\Delta\gamma$ and $\Delta\delta$, as well as their ratio. The background ratios for the CME, $\Delta\gamma_{bkg}$, were found to be less than unity, which aligns with experimental findings. Conversely, the total correlation ratios, $\Delta\gamma_{total}$, exceeded those of the background, in line with theoretical expectations. Nonetheless, the similar $\Delta\gamma$ values obtained with and without the CME indicate a very weak CME signal in isobar collisions.
The ratios associated with the CME background, $\Delta\gamma_{bkg}$, are typically below unity, which concurs with experimental observations. In contrast, the ratios of the total correlation, $\Delta\gamma_{total}$, exceed those of the background, adhering to theoretical forecasts. Nonetheless, the comparable $\Delta\gamma$ values seen in simulations both with and without the CME imply a very modest CME signal within isobar collisions.  Finally, the correlation $R_{\Psi_{2}}(\Delta S^{''})$ for Ru + Ru and Zr + Zr collisions  reveals analogous behaviors, implying a limited effect of the chiral anomaly.

\section{ACKNOWLEDGMENTS}
This work is supported in part by the National Natural Science Foundation of China (NSFC) Grant Nos. 12235016, 12221005, 12347143, 12405150, 12375121, 12205309, 12347106, 12147101, and 12325507, the Strategic Priority Research Program of Chinese Academy of Sciences under Grant No. XDB34030000 and the National Key Research and Development Program of China under Grant No. 2022YFA1604900, and the Guangdong Major Project of Basic and Applied Basic Research under Grant No. 2020B0301030008.

\bibliographystyle{IEEEtran}
\bibliography{reference.bib}
\end{document}